\documentclass[12pt,a4paper]{article}

\usepackage{graphics}
\usepackage{latexsym}
\usepackage{amsmath}
\usepackage{amssymb}
\usepackage{slashed}
\usepackage[usenames]{color}

\title{Discrepancies Between Limits and Measurements of Warm Dark Matter Properties}
\author{Bruce Hoeneisen}
\date{\small{
Universidad San Francisco de Quito, Quito, Ecuador \\
Email: bhoeneisen@usfq.edu.ec \\
22 April 2025}
}

\begin{document}
\maketitle

\begin{abstract}
\noindent
A limit on the expansion parameter $a_{h\textrm{NR}}$ at which
dark matter becomes non-relativistic
has been obtained from	the observed minimum halo mass hosting Milky Way satellites.
This limit is in disagreement with measurements.
In the present study, we attempt to understand this disagreement.
We find that the limit does not include the following
phenomena: non-linear regeneration of the power	spectrum
of density perturbations, the stripping	of galaxy halos
by neighboring galaxies, and baryons that act as 
cold dark matter. Considering these phenomena, we find that
there is no longer a significant discrepancy between the limit and
the measurements.
\end{abstract}

\noindent
\textbf{Keywords} \\
Warm Dark Matter, Galaxy, Galaxy Formation, Dwarf Galaxies

\section{Introduction}

Most non-relativistic matter in the universe is in a ``dark matter" form
that has only been ``observed" through its gravitational interaction \cite{PDG2024}.
Let us assume that this dark matter is a gas of a single particle
species of mass $m_h$. The effect of this dark matter on cosmology,
under quite general conditions,
can be described with a single
parameter $a_{h\textrm{NR}}$ \cite{Lin} to be added to the six parameters of the
standard cold dark matter $\Lambda$CDM cosmology \cite{PDG2024}.
$a_{h\textrm{NR}}$ is the characteristic universe expansion factor $a(t)$ at which
the dark matter particles become non-relativistic.
$a(t)$ is normalized so that $a(t_0) = 1$ at the present time $t_0$.
Several estimated limits on $a_{h\textrm{NR}}$ have been summarized in
Table 3 of \cite{Lin}. The limits are
\begin{equation}
a_{h\textrm{NR}} < 6 \times 10^{-5}
\label{BBN}
\end{equation}
from Big Bang Nucleosynthesis (BBN) observations,            
\begin{equation}
a_{h\textrm{NR}} \lesssim 7 \times 10^{-6}
\label{CMB}
\end{equation}
from observed Cosmic Microwave Background (CMB) radiation fluctuations,
\begin{equation}
a_{h\textrm{NR}} \lesssim 7 \times 10^{-7}
\label{Ly-a}
\end{equation}
from observations of the Lyman-$\alpha$	forest of quasar light, and
\begin{equation}
a_{h\textrm{NR}} \lesssim 6 \times 10^{-8}
\label{small-halo}
\end{equation}
from the observed population of small galaxy halos.
In contrast, $a_{h\textrm{NR}}$ has been \textit{measured}
with dwarf galaxy rotation curves \cite{dwarf} \cite{summary}:
\begin{equation}
a_{h\textrm{NR}} = (1.39 \pm 0.24) \times 10^{-6}.
\label{dwarf}
\end{equation}
There is a discrepancy between (\ref{small-halo}) and (\ref{dwarf}).
Each of these measurements and limits has its own issues.
The purpose of the present study is to try to understand
the discrepancy between (\ref{small-halo}) and (\ref{dwarf}).

The general conditions mentioned above assume dark matter
with or without interactions with the Standard Model sector
or self-interactions, as long as the chemical potential is non-positive.
In the present study, we further assume that dark matter was once
in thermal equilibrium with the early Standard Model sector, \textit{i.e.}
we consider ``thermal relic" dark matter. 

The limit (\ref{small-halo}) is obtained from the minimum halo mass hosting Milky Way satellites
\cite{Lin},
\cite{Lin2}- 
\cite{Jethwa}:
\begin{equation}
M_h < 5.4 \times 10^8 \textrm{ M}_\odot,
\label{Mh}
\end{equation}
at $2 \sigma$ confidence \cite{Lin}.
The corresponding observed stellar mass is of order $M_* \approx 10^5$ M$_\odot$ \cite{Lin}
(that agrees with the calculation in \cite{Weinberg}).
The authors of \cite{Lin} state that (\ref{small-halo}) is an \textit{estimate},
and, due to non-linearities, 
``should be more rigorously studied by cosmological simulation".
Below we fill in this gap.

The present work is a continuation of \cite{understanding} and \cite{extended}.
These references may be consulted for details.
To make the present article self-contained we list some definitions in 
Section \ref{definitions}. Warm dark matter introduces 
the ``free-streaming cut-off" discussed in Section \ref{PS}.
Numerical integration of hydrodynamical equations is presented in Section \ref{vel_disp}.
The non-linear regeneration of small scale structure is discussed in Section \ref{discussion}, 
the bottom-up and top-down evolution of structure in the warm dark matter
scenario is presented in Section \ref{up_down}, and the effect of baryons is
discussed in Section \ref{baryons}.
Conclusions follow.

\section{Definitions}
\label{definitions}

We use the notation and parameter values of \cite{PDG2024}.
Our definitions follow.
Let $v_{h\textrm{rms}}(a)$ be the 3-D
root-mean-square thermal velocity of non-relativistic warm dark matter particles
at expansion parameter $a(t)$ when the universe is nearly homogeneous. 
$v_{h\textrm{rms}}(a)$ scales as $a^{-1}$, so
\begin{equation}
v_{h\textrm{rms}}(1) = a \cdot v_{h\textrm{rms}}(a)
\label{vhrms1}
\end{equation}
is an adiabatic invariant, and
\begin{equation}
a_{h\textrm{NR}} \equiv 1.03 \frac{v_{h\textrm{rms}}(1)}{c},
\label{ahNR3}
\end{equation}
where 1.03 is a threshold factor for bosons (0.98 for fermions) \cite{summary}.
Warm dark matter free-streaming attenuates the \textit{linear} 
power spectrum of density perturbations $P_\textrm{CDM}(k)$
of the $\Lambda$CDM cosmology, 
at large comoving wavevector $k$, by a factor \cite{Viel}
\begin{equation}
\tau^2(k) \equiv \frac{P_\textrm{WDM}(k)}{P_\textrm{CDM}(k)} 
= \left[ 1 + (\alpha k)^{2 \nu} \right]^{-10/\nu},
\label{tau2}
\end{equation}
where $\nu = 1.12$, and
\begin{equation}
\alpha = 0.049 \left( \frac{m_h}{1 \textrm{keV}} \right)^{-1.11} \left( \frac{\Omega_c}{0.25} \right)^{0.11}
\left( \frac{h}{0.7} \right)^{1.22} \frac{1}{h} \textrm{ Mpc}.
\label{alpha}
\end{equation}
These equations, often used in the literature, \textit{define} the ``standard thermal relic mass" $m_h$.
Limits in the literature often obtain a lower bound to this $m_h$.
The actual dark matter particle mass is model-dependent, see Table 4 of \cite{summary}.
Equation (\ref{tau2}) can be approximated by
\begin{equation}
\tau^2(k) \approx \exp{\left( - \frac{k^2}{k_\textrm{fs}^2} \right)}, 
\qquad \textrm{where} \qquad k_\textrm{fs} = \frac{1}{2.59 \cdot \alpha},
\label{tau22}
\end{equation}
except for a ``tail" to be discussed below.
$k_\textrm{fs}$ is the comoving free-streaming wavevector.
At this wavevector, $\tau^2(k_\textrm{fs}) = 1/e$ (other definitions in the literature
are 1/2 or 1/4).
An effective warm dark matter comoving free-streaming length can be defined as
\begin{equation}
l_\textrm{fs} = 0.26 \frac{2 \pi}{k_\textrm{fs}},
\label{lfs}
\end{equation}
where the factor 0.26 comes from an integration over angles.
$l_\textrm{fs}$ is approximately proportional to $v_{h\textrm{rms}}(1)$:
\begin{equation}
l_\textrm{fs} \approx 4.9 \left[ \frac{\textrm{Mpc}}{\textrm{km/s}} \right] \cdot v_{h\textrm{rms}}(1).
\end{equation}
An analytic expression that relates directly $v_{h\textrm{rms}}(1)$ and $k_\textrm{fs}$ 
(that does not include free-streaming during radiation domination while inside the horizon) 
is \cite{Boyanovsky} \cite{Piattella}
\begin{equation}
k_\textrm{fs}(t_\textrm{eq}) = 
  \frac{1.455}{\sqrt{2}} \sqrt{\frac{4 \pi G \Omega_m \rho_\textrm{crit} a_\textrm{eq}}{v^2_{h\textrm{rms}}(1)}}.
\label{kfs}
\end{equation}
This comoving wavevector approximately separates growing from decaying modes at the time $t_\textrm{eq}$
of equal radiation and matter densities. Thereafter, $k_\textrm{fs}$ grows as $\propto a^{1/2}$,
allowing regeneration of small scale structure, giving $\tau(k)$ a ``tail"
that depends on $a(t)$.
The 3-D Fourier transform of (\ref{tau22}) defines a linear free-streaming mass scale
\begin{equation}
M_\textrm{fs} \approx \frac{4}{3} \pi \left( \frac{1.555}{k_\textrm{fs}} \right)^3 \Omega_m \rho_\textrm{crit},
\label{Mfs}
\end{equation}
where $\Omega_m \rho_\textrm{crit}$ is the present mean matter density of the universe.
In conclusion, each of the observables $m_h$, $k_\textrm{fs}$, $l_\textrm{fs}$, $v_{h\textrm{rms}}(1)$, 
$a_{h\textrm{NR}}$ and $M_\textrm{fs}$ determines all the others.

Astronomers obtain the galaxy ``stellar mass" $M_*$ 
and ``halo mass" $M_h$ (also called ``virial mass", which is a missnomer
in view of \cite{Lelli}). The ``halo mass"  $M_h$ is often defined as the galaxy
mass contained inside a radius $r_{200}$ at which the dark matter density reaches
200 times the mean dark matter density of the universe at the observed redshift
of the galaxy \cite{Munshi}. The galaxy stellar mass $M_*$ is obtained from measurements of
the relative luminosities of the galaxy with several filters, the measurement
of its redshift, and from stellar synthesis models. The galaxy halo mass 
$M_h$ is \textit{estimated} from $M_*$, stellar images,
(warm) dark matter simulations and abundance matching \cite{Munshi}, 
or exceptionally, with gravitational lensing measurements \cite{Lelli}.

\section{The linear theory}
\label{PS}

The Press-Schechter prediction \cite{PS}, or its Sheth-Tormen extensions
\cite{Sheth_Tormen} \cite{Sheth_Mo_Tormen},
depend on the variance of
the relative density perturbation $\delta(\textbf{x}) \equiv (\rho(\textbf{x}) - \bar{\rho})/\bar{\rho}$
on the linear total (dark matter plus baryon) mass scale $M_\textrm{PS}$, at redshift $z$ \cite{Weinberg} \cite{UV}:
\begin{equation}
\sigma^2(M_\textrm{PS}, z, k_\textrm{fs}) = \frac{f^2}{(2 \pi)^3 (1 + z)^2} 
\int_0^\infty 4 \pi k^2 dk P_{\textrm{CDM}}(k) \tau^2(k) W^2(k),
\end{equation}
and so depends on the free-streaming cut-off factor $\tau^2(k)$, and on the
window function $W(k)$ that defines the linear mass scale $M_\textrm{PS}$.
For a Gaussian window function
\begin{equation}
W(k) = \exp{\left( -\frac{k^2}{2 k_0^2} \right)}, \qquad \textrm{and} \qquad
M_\textrm{PS} = \frac{4}{3} \pi \left( \frac{1.555}{k_0} \right)^3 \Omega_m \rho_\textrm{crit}.
\label{Wk}
\end{equation}
We calibrate the amplitude of $P_{\textrm{CDM}}(k)$ with $\sigma_8 = 0.811$ \cite{PDG2024}
with a ``top-hat" window function of radius $8/h$ Mpc, and $f = 1$.
For our examples of Section \ref{vel_disp}, we take $f = 1/0.79$ 
due to the recent accelerated expansion of the universe \cite{Weinberg}.
The resulting predictions of the galaxy
stellar mass distributions with $M_\textrm{PS} \approx 10^{1.5} M_*$, and the galaxy ultra-violet luminosity
distributions, are excellent in a wide range of redshifts $z$
(provided $\tau(k)$ acquires a ``tail" discussed in Section 6 below),
see, for example, \cite{JWST}. 

\begin{figure}
\begin{center}
\scalebox{0.3}
{\includegraphics{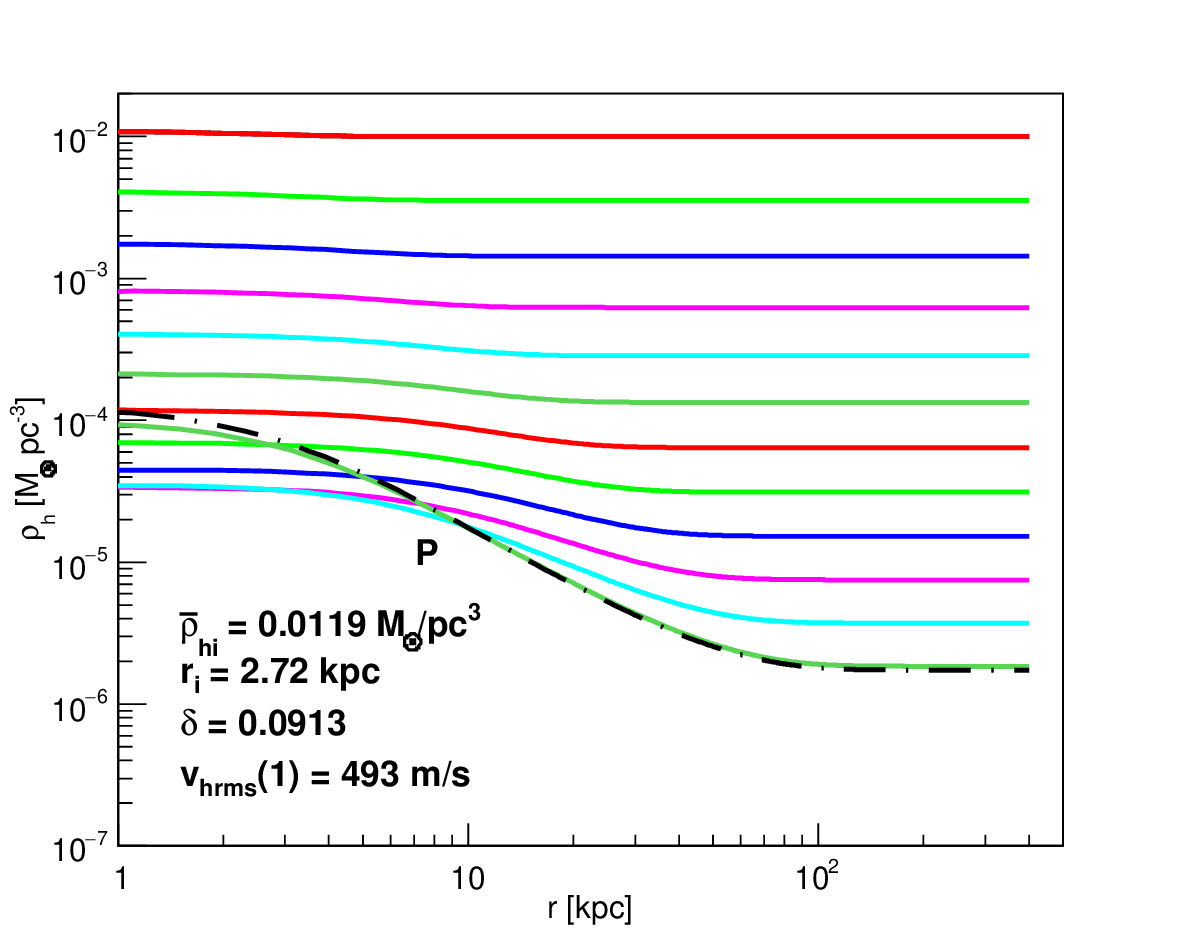}}
\scalebox{0.3}
{\includegraphics{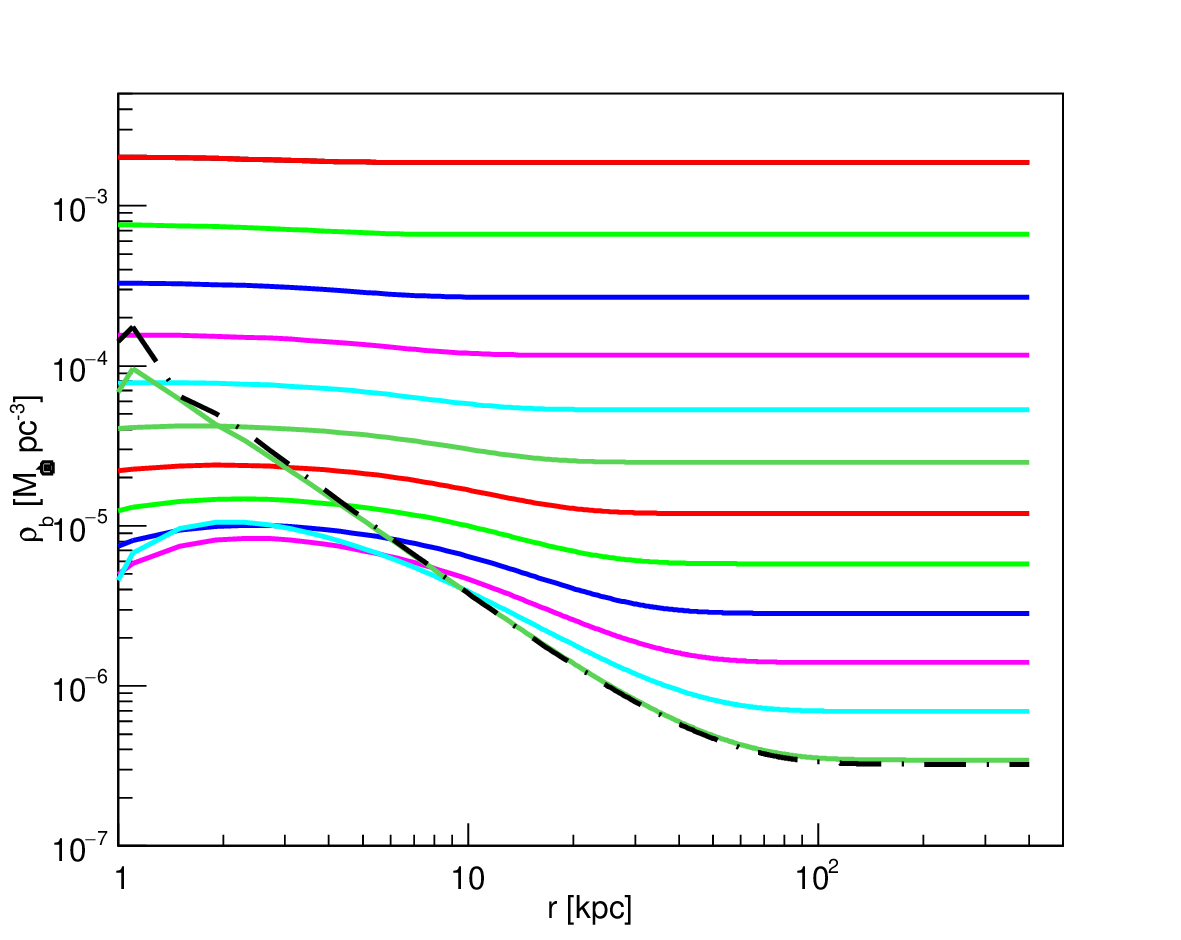}}
\scalebox{0.3}
{\includegraphics{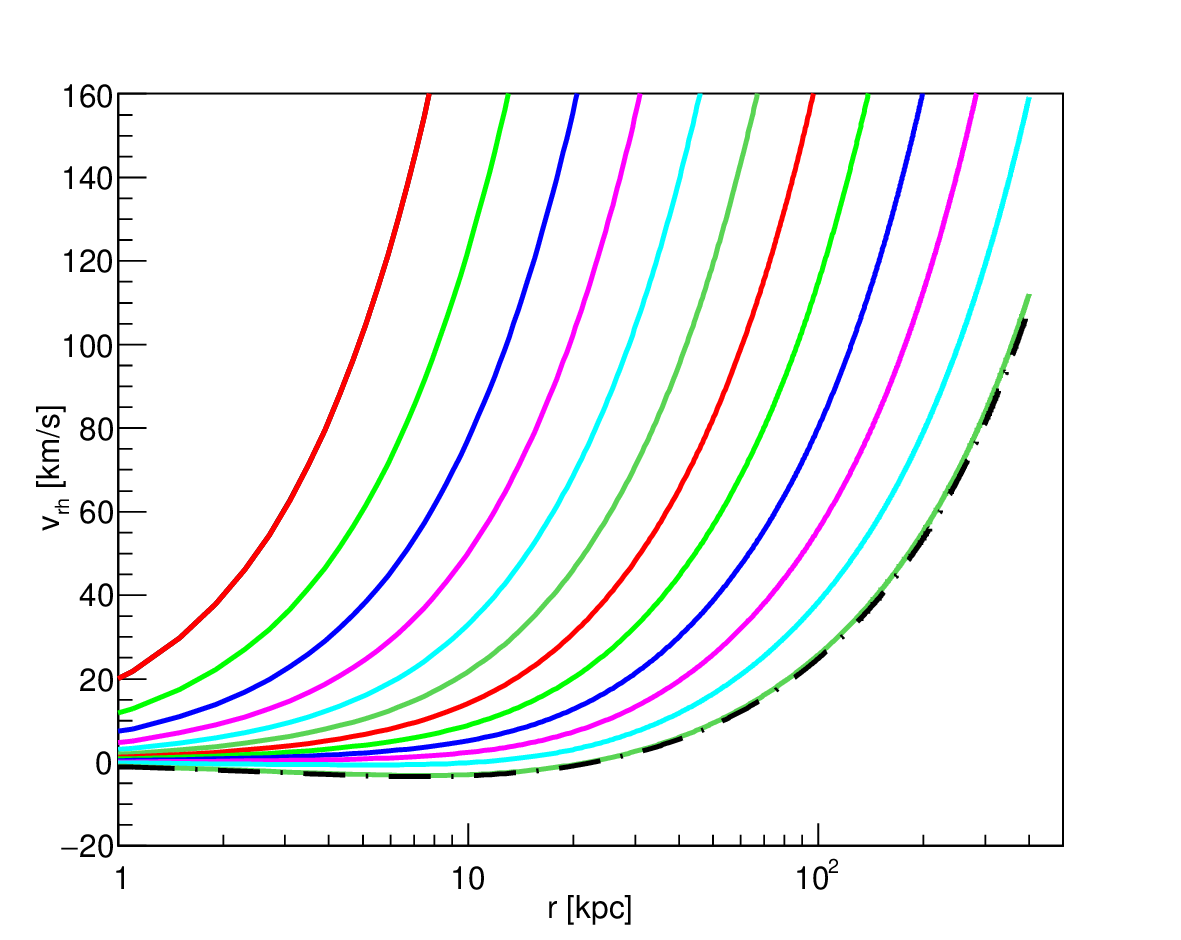}}
\scalebox{0.3}
{\includegraphics{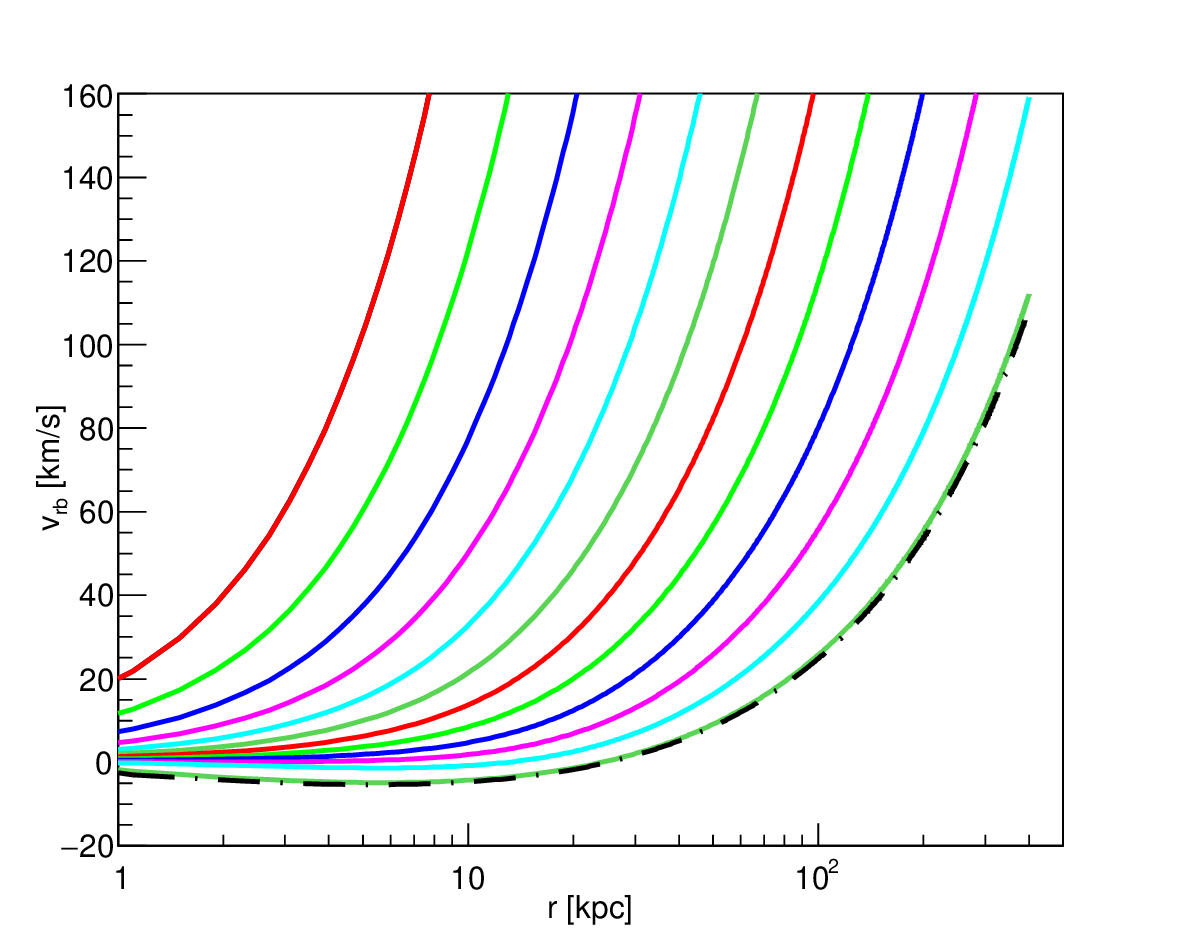}}
\caption{
The dark matter and baryon densities and radial velocities,
are shown as a 
function of the proper radius $r$, at times
that increase by factors $\sqrt{2}$ (except for the last
dashed line).
The initial redshift of the numerical integration is $z_i = 65.9$.
The parameters of this simulation are
$\bar{\rho}_{hi}=0.0119$ M$_\odot/$pc$^3$,
proper $r_i = 2.72$ kpc,
$\delta=0.0913$, and
$v_{h\textrm{rms}}(1)=493$ m/s.
The pivot point P has $r_{h\textrm{P}} = 9$ kpc and 
$\rho_{h\textrm{P}} = 2 \times 10^{-5}$ M$_\odot/$pc$^3$. 
$M_\textrm{PS} = 10^9$ M$_\odot$. $M_h = 3 \times 10^8$ M$_\odot$.
$\sigma(M_\textrm{PS}, z_i, k_\textrm{fs}) = 0.038$.
$V_\textrm{rot} = 9.4$ km/s.
}
\label{10_9}
\end{center}
\end{figure}

\section{Galaxy formation}
\label{vel_disp}

The formation of a galaxy with warm dark matter and baryons 
is described by hydrodynamical equations \cite{understanding}.
These equations are valid for collisionless, or collisional
warm dark matter particles, so long as collisions are elastic.
A numerical integration of these hydrodynamical equations is
presented in Figure \ref{10_9}, assuming spherical symmetry. 
We are interested in the limit of very small dark matter particle
mass and large velocity dispersion, so, for all
examples in this article, we choose
\begin{equation}
m_h = 0.15 \textrm{ keV}, \qquad  \textrm{and} \qquad v_{h\textrm{rms}}(1) = 493 \textrm{ m/s}.
\label{mh_vhrms1}
\end{equation}
This particular choice of parameters is justified in Table 4 of \cite{summary}.
According to (\ref{ahNR3}), (\ref{alpha}) and (\ref{lfs}) the comoving 
free-streaming cut-off
wavevector is $k_\textrm{fs} \approx 0.67$ Mpc$^{-1}$, the comoving effective
free-streaming length is $l_\textrm{fs} \approx 2.4$ Mpc,
and the dark matter becomes non-relativistic at expansion
parameter $a_{h\textrm{NR}} \approx 1.7 \times 10^{-6}$.
Note that this $a_{h\textrm{NR}}$ is in agreement with
the measurement (\ref{dwarf}), and in disagreement with the limit (\ref{small-halo}).
The linear total (dark matter plus baryon) mass corresponding
to $k_\textrm{fs}$ is $M_\textrm{fs} \approx 2 \times 10^{12}$ M$_\odot$.

The numerical integration in Figure \ref{10_9} begins at 
redshift $z_i = 65.9$, with initial dark matter and baryon densities
\begin{equation}
\rho_{hi}(r) = \frac{\Omega_h}{\Omega_b} \rho_{bi}(r) =
\left< \bar{\rho}_{hi} \right> \left[ 1 + \delta \exp{(-r^2 / r_i^2)}\right],
\label{rho_i}
\end{equation}
with $\left< \bar{\rho}_{hi} \right> = 0.0119$ M$_\odot/$pc$^3$
(corresponding to $z_i$), $r_i = 2.72$ kpc
and $\delta = 0.0913$. The initial velocities are
$v_{hi}(r) = v_{bi}(r) = H(t_i) r$ (and a tiny correction due to $\delta$).
The linear ``Press-Schechter" mass corresponding to $r_i$ is
$M_\textrm{PS} = 10^{9}$ M$_\odot$, and the relative mass fluctuation
on this mass scale at $z_i$ is $\sigma = 0.038$.
The subscript $h$ stands for dark matter, and the subscript $b$ stands
for ``baryons".

A moment past the last time
presented in Figure \ref{10_9}, the density $\rho_h(r)$ becomes
a ``cored isothermal sphere" \cite{understanding} \cite{extended},
\textit{i.e.} a solution of the static limit of the hydrodynamical equations,
with a density run $\rho_h(r) \propto r^{-2}$ at $r \gg r_c$, and a core radius
\begin{equation}
r_c = \sqrt{\frac{\left< v_{hr}^2 \right>}{2 \pi G \rho_{hc}}},
\label{rc}
\end{equation}
with the root-mean-square dark matter radial velocity 
\begin{equation}
\sqrt{\left< v_{hr}^2 \right>} \approx \frac{v_{h\textrm{rms}}(1)}{\sqrt{3}}
\left( \frac{\rho_{hc}}{\Omega_c \rho_\textrm{crit}} \right)^{1/3},
\label{vhr}
\end{equation}
that is independent of $r$ and $t$ \cite{extended}.
The radius of the isothermal sphere keeps growing due to the
expansion of the universe \cite{extended}.

The numerical integration of Figure \ref{10_9} obtains
the ``pivot point" P with $r_{h\textrm{P}} = 9$ kpc and
$\rho_{h\textrm{P}} = 2 \times 10^{-5}$ M$_\odot/$pc$^3$.
From these numbers, we obtain the radius at which the
dark matter halo density is 200 times the present mean
dark matter density of the universe (since satellites of the Milky Way are
observed at $z \approx 0$): $r_{200} = 16$ kpc, and
the corresponding halo mass 
\begin{equation}
M_h \equiv M_{200} = \frac{2 \left< v_{hr}^2 \right>}{G} r_{200} = 3 \times 10^8 \textrm{ M}_\odot.
\label{M200} 
\end{equation}
This is the mass that needs to be compared to the minimum observed halo mass
(\ref{Mh}) \cite{Munshi}. 
Note that
\begin{equation}
\rho_h(r) r^2 = \rho_{hc} r_c^2 = \rho_{h\textrm{P}} r_{h\textrm{P}}^2 = 200 \Omega_c \rho_\textrm{crit} r_{200}^2,
\end{equation}
and
\begin{equation}
\sqrt{\left< v_{hr}^2 \right>} = \sqrt{2 \pi G \rho_{h\textrm{P}} r_{h\textrm{P}}^2}.
\end{equation}

\begin{figure}
\begin{center}
\scalebox{0.3}
{\includegraphics{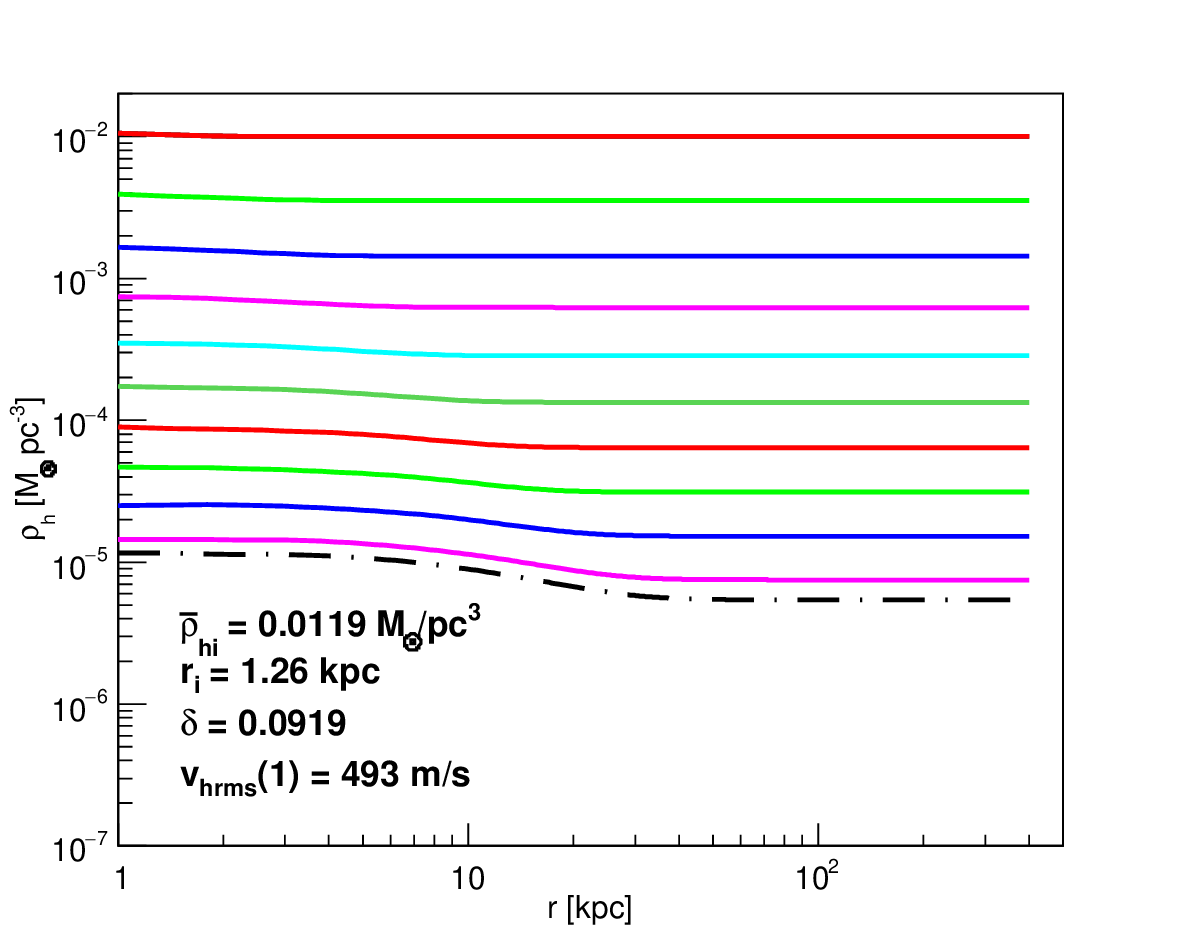}}
\scalebox{0.3}
{\includegraphics{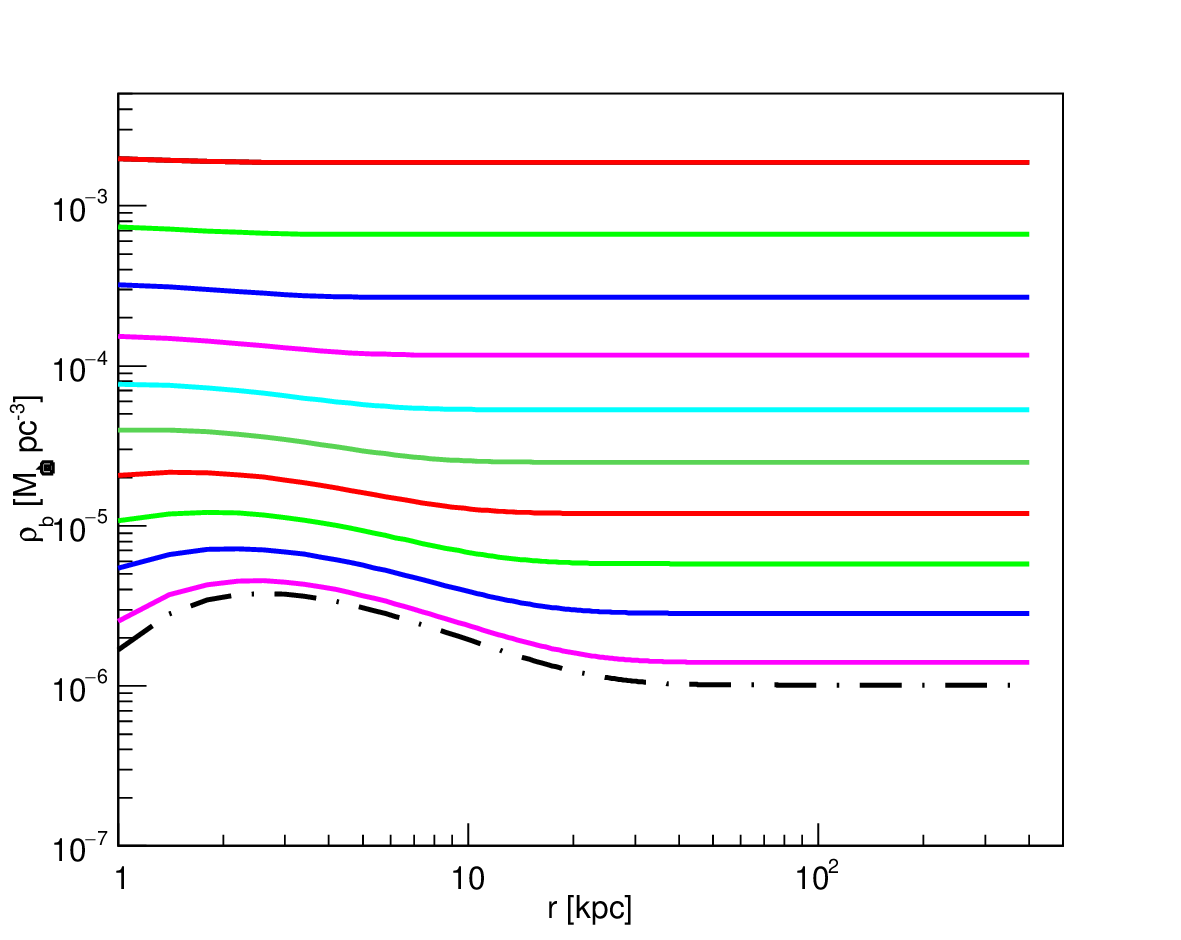}}
\scalebox{0.3}
{\includegraphics{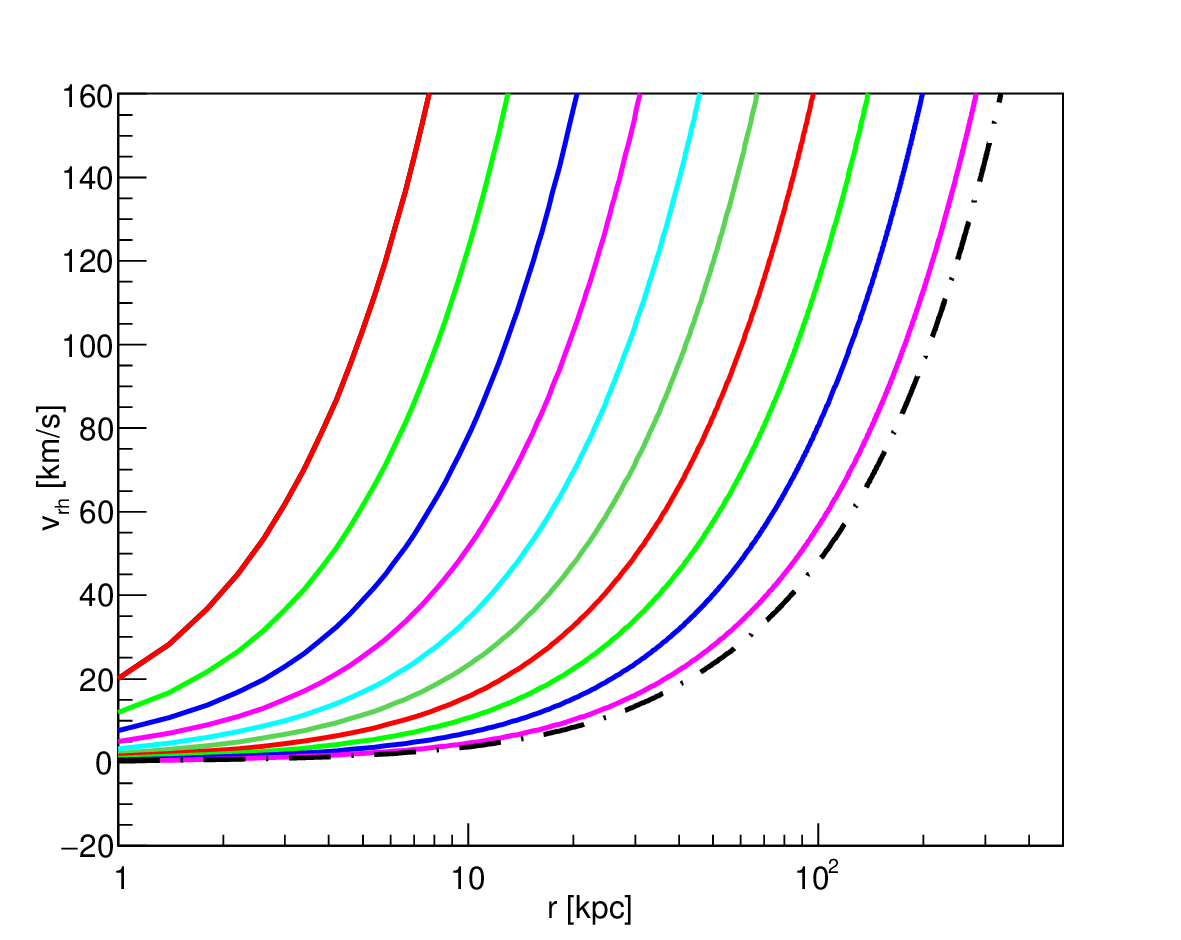}}
\scalebox{0.3}
{\includegraphics{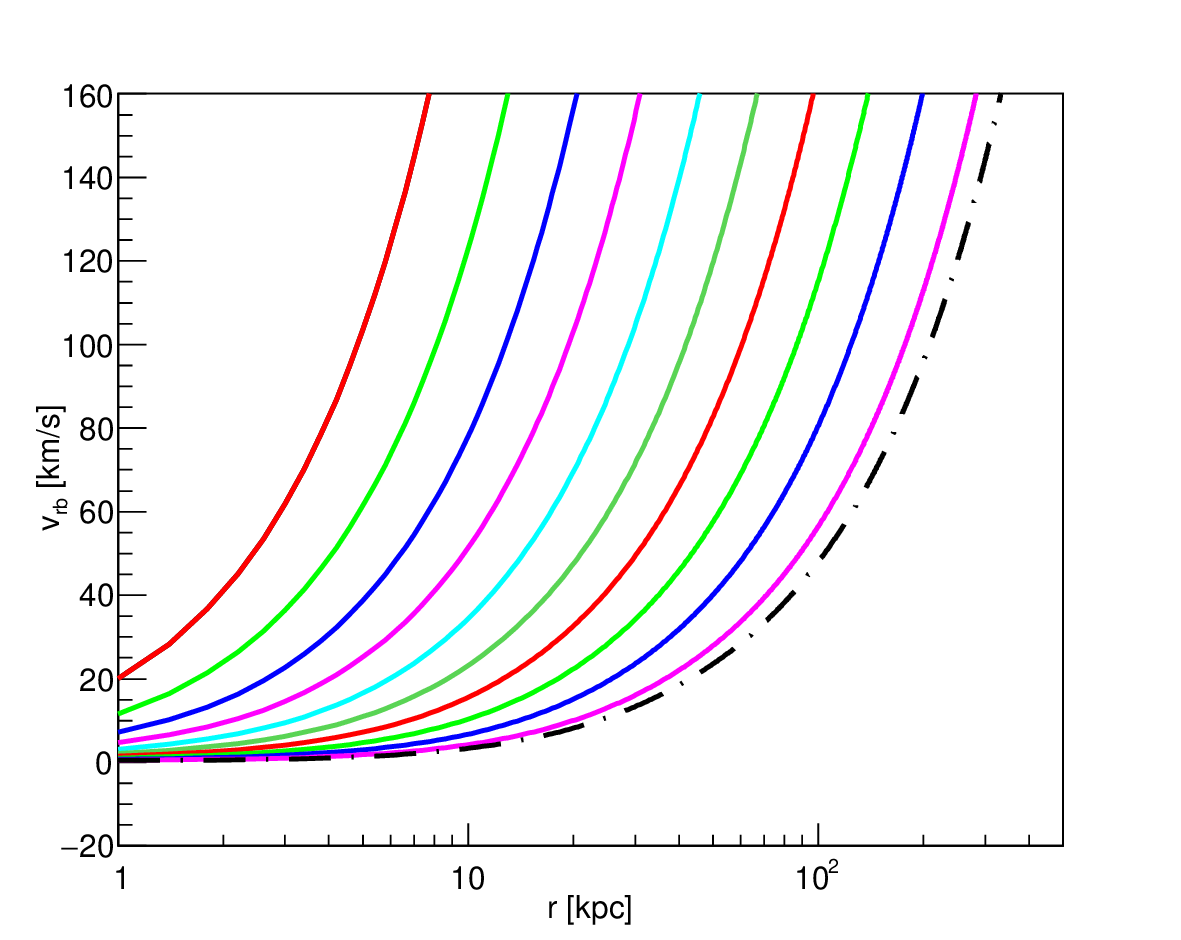}}
\caption{
The dark matter and baryon densities and radial velocities
are shown as a 
function of the proper radius $r$, at times
that increase by factors $\sqrt{2}$
(except the last dashed curve).
The initial redshift of the numerical integration is $z_i = 65.9$.
The parameters of this simulation are
$\bar{\rho}_{hi}=0.0119$ M$_\odot/$pc$^3$,
$r_i = 1.26$ kpc,
$\delta=0.092$, and
$v_{h\textrm{rms}}(1)=493$ m/s.
There is no pivot point P, so no galaxy halo forms.
$M_\textrm{PS} = 10^8$ M$_\odot$. 
$\sigma(M_\textrm{PS}, z_i, k_\textrm{fs}) = 0.038$.
}
\label{10_8}
\end{center}
\end{figure}

\begin{figure}
\begin{center}
\scalebox{0.3}
{\includegraphics{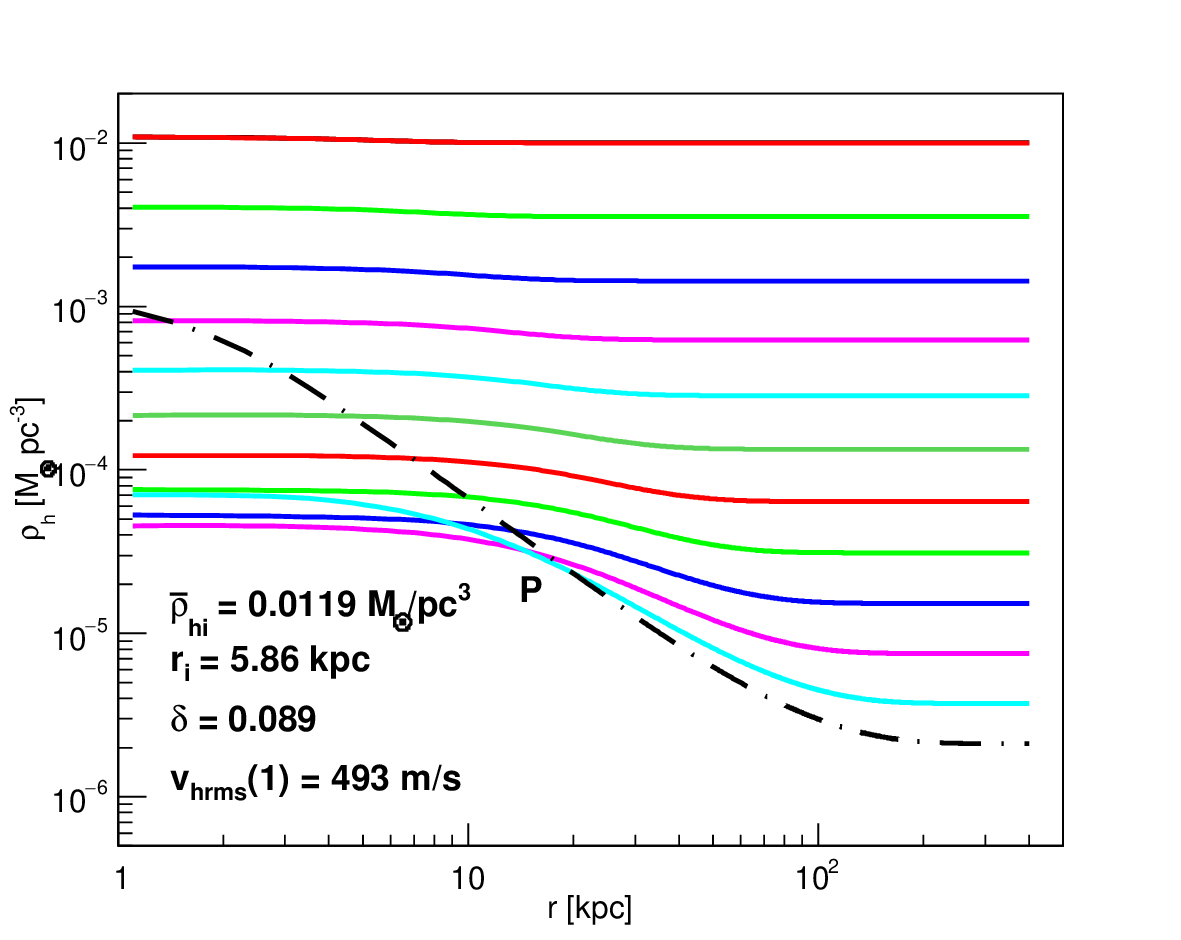}}
\scalebox{0.3}
{\includegraphics{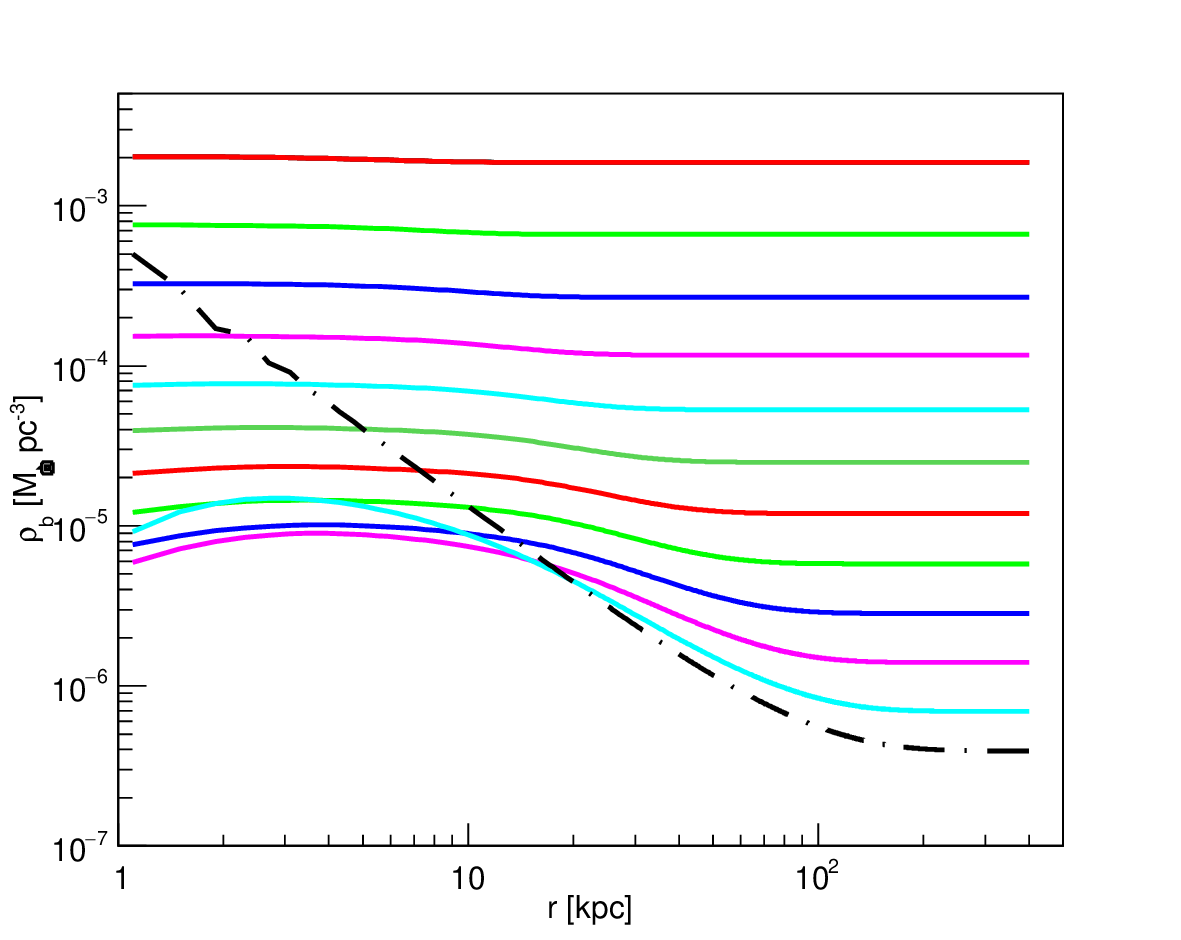}}
\scalebox{0.3}
{\includegraphics{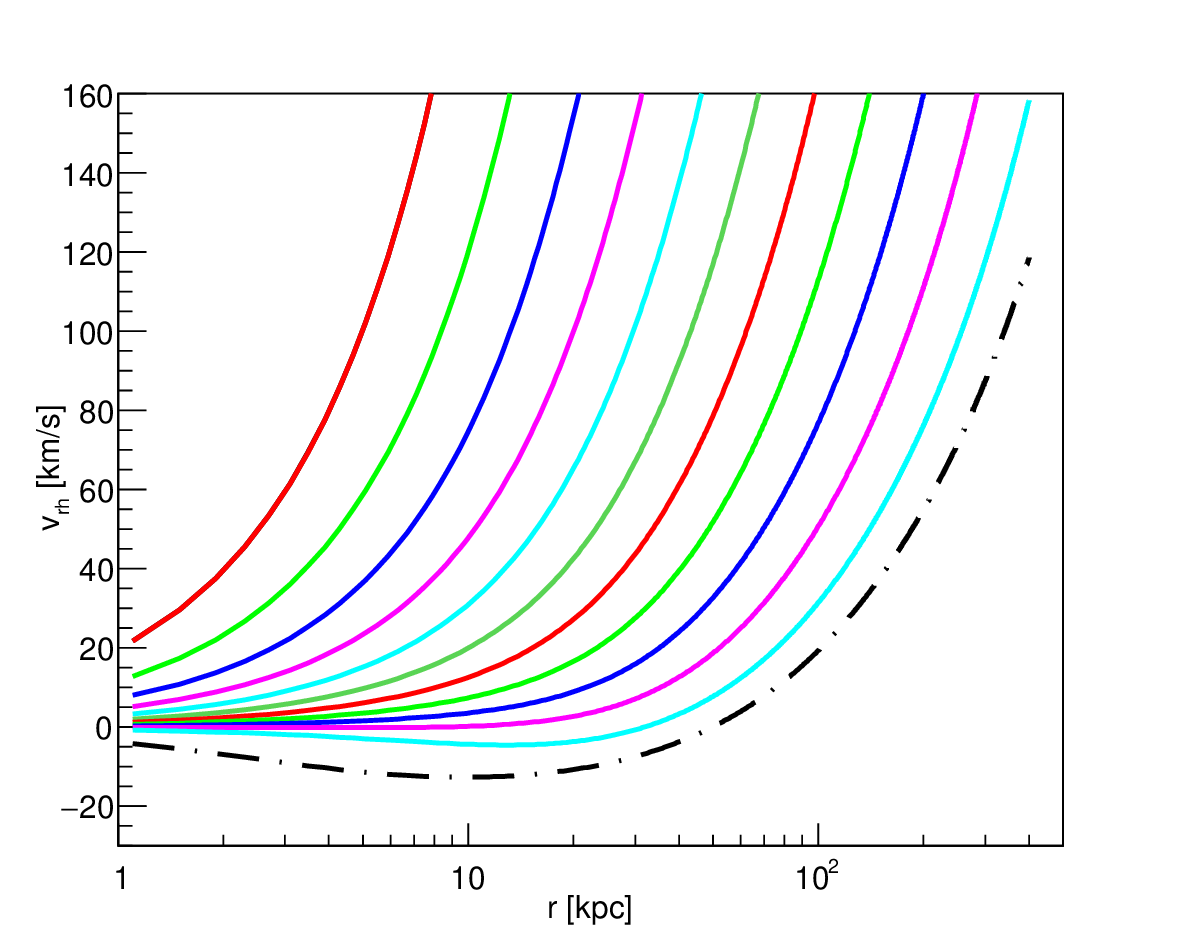}}
\scalebox{0.3}
{\includegraphics{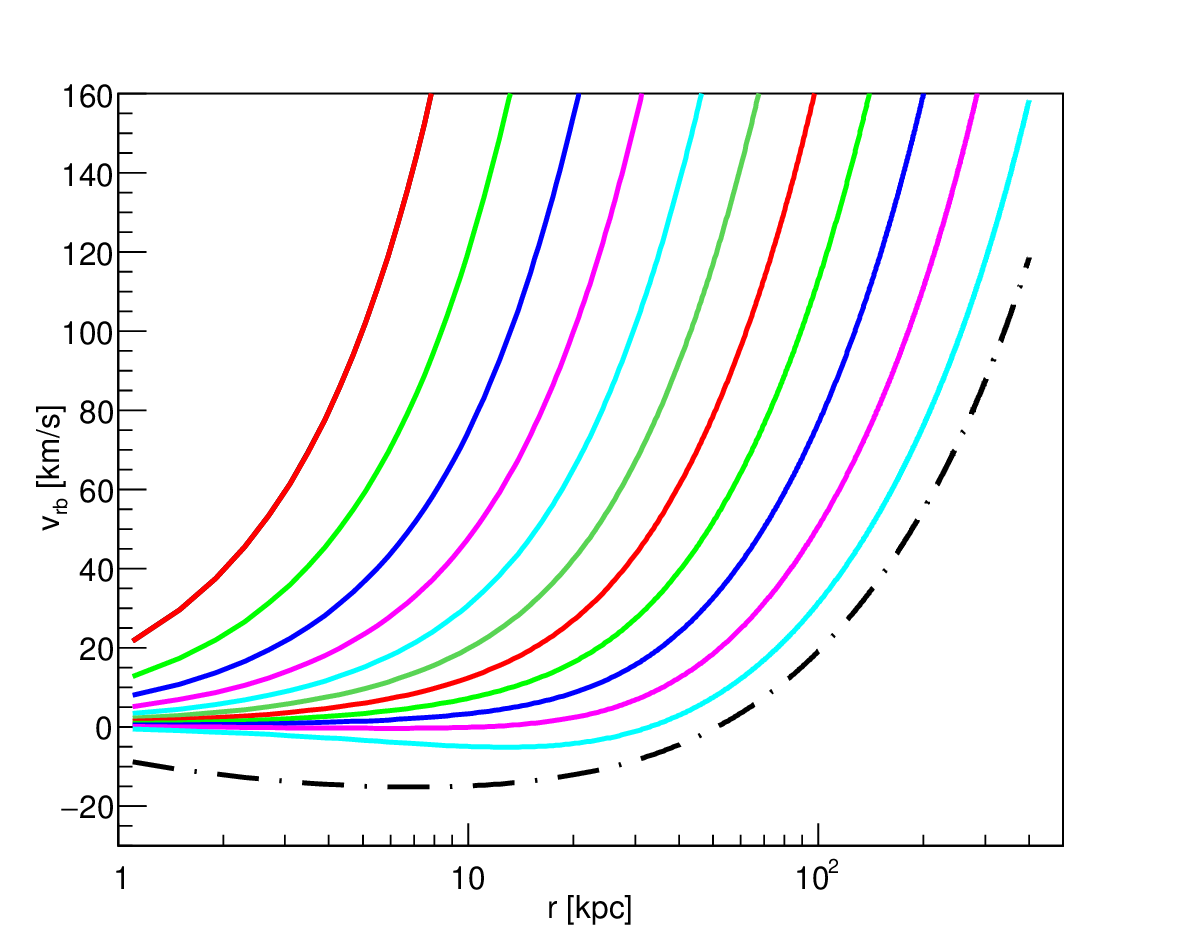}}
\caption{
The dark matter and baryon densities and radial velocities
are shown as a 
function of the proper radius $r$, at times
that increase by factors $\sqrt{2}$ (except for the last
dashed line).
The initial redshift of the numerical integration is $z_i = 65.9$.
The parameters of this simulation are
$\bar{\rho}_{hi}=0.0119$ M$_\odot/$pc$^3$,
$r_i = 5.86$ kpc,
$\delta=0.089$, and
$v_{h\textrm{rms}}(1)=493$ m/s.
The pivot point P has $r_{h\textrm{P}} = 17$ kpc and 
$\rho_{h\textrm{P}} = 2.3 \times 10^{-5}$ M$_\odot/$pc$^3$. 
$M_\textrm{PS} = 10^{10}$ M$_\odot$. $M_h = 3 \times 10^9$ M$_\odot$.
$\sigma(M_\textrm{PS}, z_i, k_\textrm{fs}) = 0.037$.
$V_\textrm{rot} = 19$ km/s.
}
\label{10_10}
\end{center}
\end{figure}

\begin{figure}
\begin{center}
\scalebox{0.3}
{\includegraphics{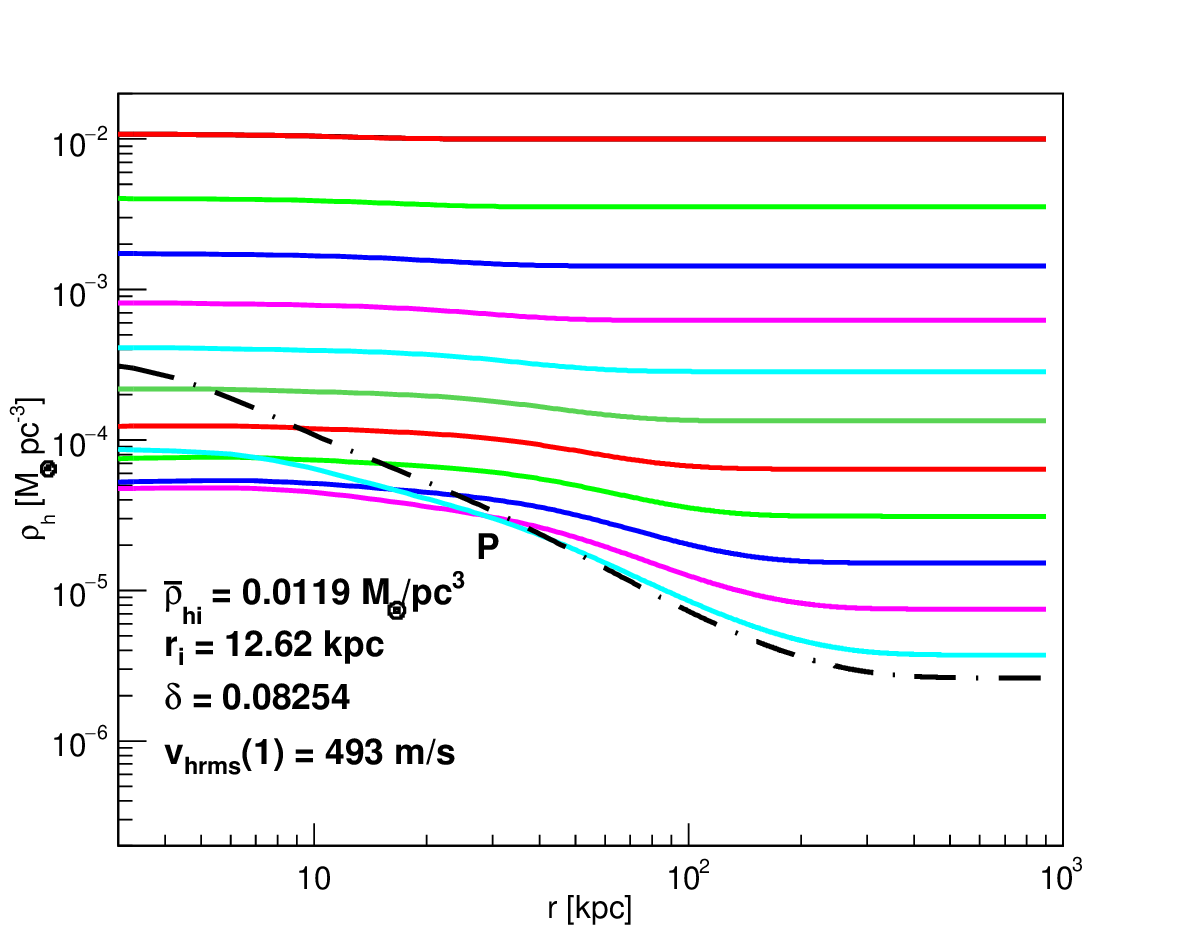}}
\scalebox{0.3}
{\includegraphics{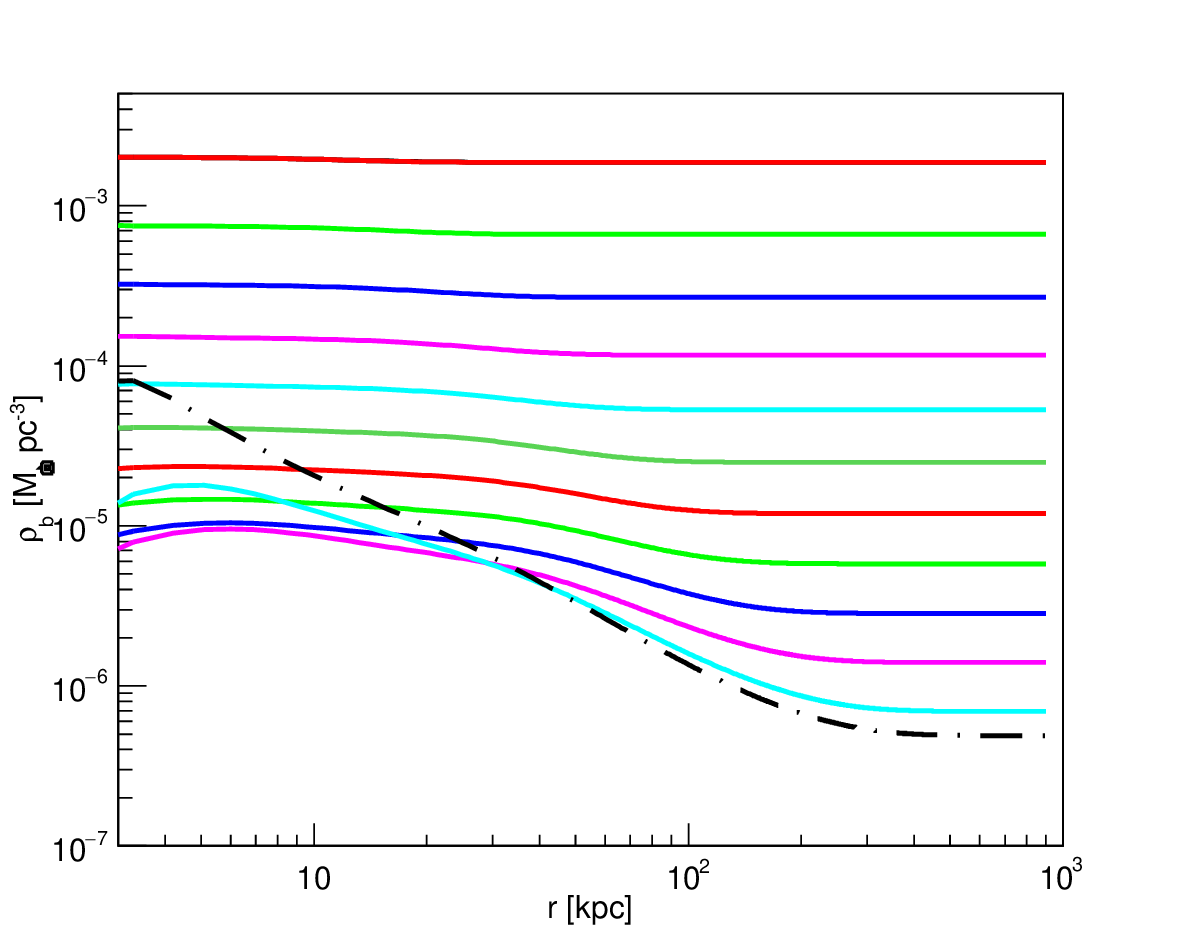}}
\scalebox{0.3}
{\includegraphics{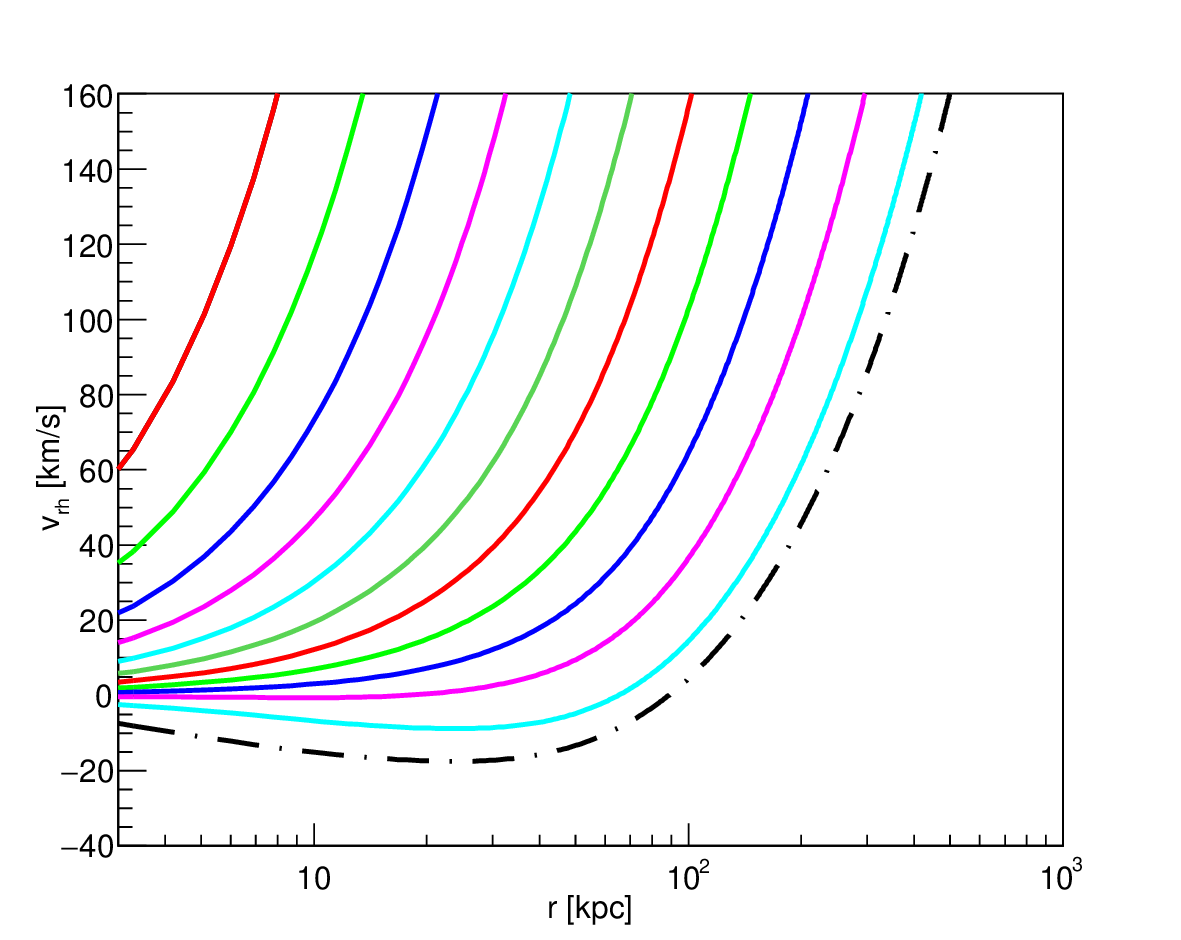}}
\scalebox{0.3}
{\includegraphics{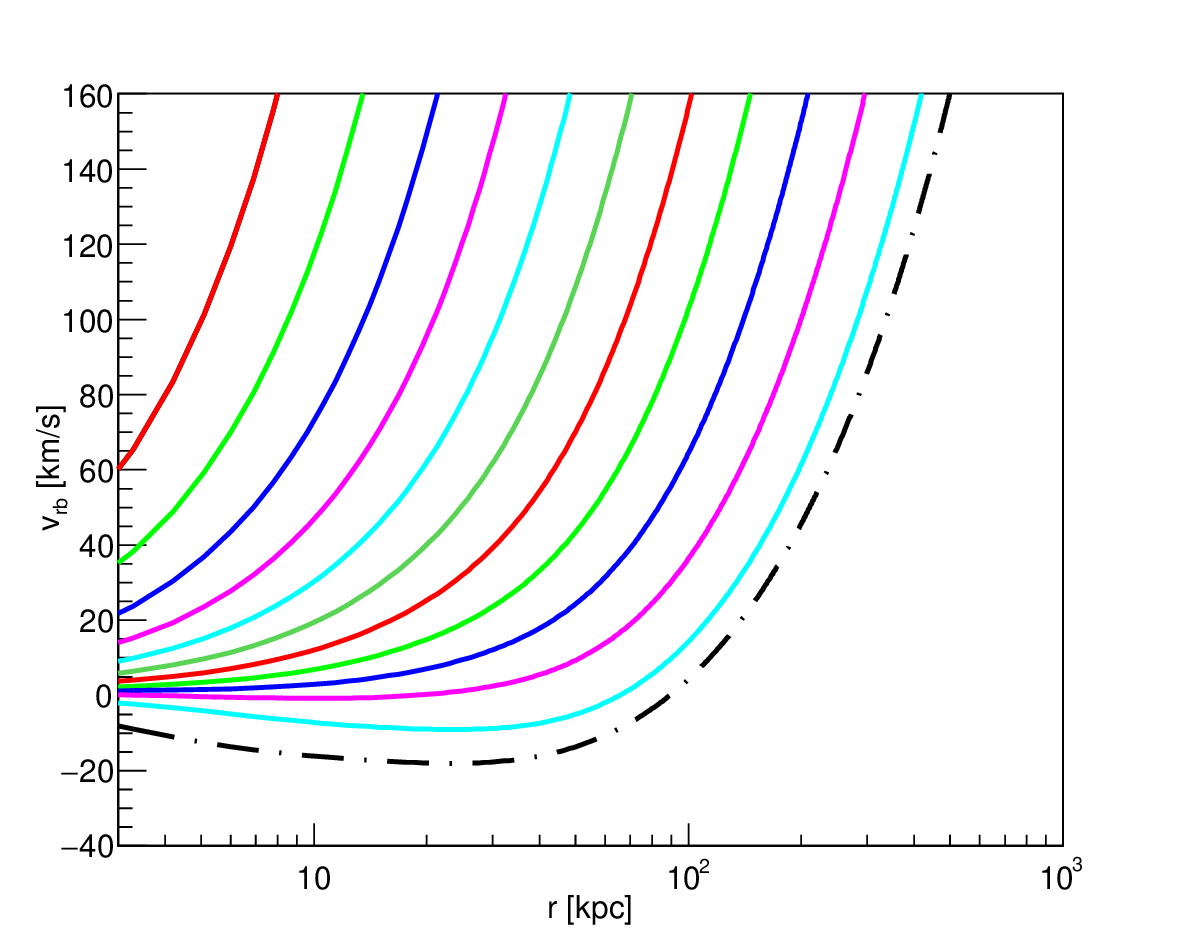}}
\caption{
The dark matter and baryon densities and radial velocities
are shown as a 
function of the proper radius $r$, at times
that increase by factors $\sqrt{2}$ (except the last dashed line).
The initial redshift of the numerical integration is $z_i = 65.9$.
The parameters of this simulation are
$\bar{\rho}_{hi}=0.0119$ M$_\odot/$pc$^3$,
$r_i = 12.625$ kpc,
$\delta=0.08254$, and
$v_{h\textrm{rms}}(1)=493$ m/s.
The pivot point P has $r_{h\textrm{P}} = 35$ kpc and 
$\rho_{h\textrm{P}} = 2.5 \times 10^{-5}$ M$_\odot/$pc$^3$. 
$M_\textrm{PS} = 10^{11}$ M$_\odot$. $M_h = 3 \times 10^{10}$ M$_\odot$.
$\sigma(M_\textrm{PS}, z_i, k_\textrm{fs}) = 0.036$.
$V_\textrm{rot} = 41$ km/s.
}
\label{10_11}
\end{center}
\end{figure}

The origin of the core is explained in \cite{extended}, and is of 
cosmological origin, see (\ref{vhr}).
In the example of Figure \ref{10_9}, the core radius and density are
$r_c = 2.0$ kpc and $\rho_{hc} = 4 \times 10^{-4}$ M$_\odot/$pc$^3$.
$\sqrt{\left< v_{hr}^2 \right>} = 6.6$ km/s, so the velocity of rotation
of a test particle is $V_\textrm{rot} = \sqrt{2 \left< v_{hr}^2 \right>} = 9.4$ km/s.
This is a ``small" galaxy indeed!

The simulation for $M_\textrm{PS} = 10^{8}$ M$_\odot$,
with the listed parameters, does not form a halo, see Figure \ref{10_8}.
Simulations of more massive galaxies are presented in 
Figures \ref{10_10} and \ref{10_11}. The halo masses
$M_h$ are given in the figure captions.
For each simulation, the value of $\sigma(M_\textrm{PS}, z, k_\textrm{fs})$
(with a top-hat window function)
is also given in the figure caption.

\section{Non-linear regeneration of $P(k)$}
\label{discussion}

Let us consider Figure \ref{10_9}.
How probable is the density fluctuation (\ref{rho_i}) at $z_i = 65.9$?
This density fluctuation has a Fourier transform (\ref{Wk})
with a cut-off comoving wavevector $k_0 = 1.555/(2.72 \textrm{ kpc} \cdot 65.9) = 8.6$ Mpc$^{-1}$.
For our example (\ref{mh_vhrms1}), $k_\textrm{fs} = 0.67$ Mpc$^{-1}$.
The \textit{linear} $P(k)$ is exponentially attenuated for
$k > k_\textrm{fs}$, so the density fluctuation in
Figure 1 is improbable. This is the argument in \cite{Lin}.

However, there are at least four phenomena not considered in \cite{Lin}:
\begin{enumerate}
\item
The linear power spectrum $P(k)$ of (\ref{tau22}) develops a non-linear
regenerated tail.
\item
A majority of ``small" galaxies have lost matter to their neighbors. 
\item
After decoupling, baryons behave as cold dark matter.
\item
If dark matter particles are bosons that decouple while ultra-relativistic,
\textit{i.e.} are in thermal equilibrium with
zero chemical potential while ultra-relativistic,
then the excess of low momentum particles behaves as cold dark matter,
see figure 10 of \cite{Boyanovsky}.
\end{enumerate}

The non-linear regeneration of $P(k)$ for warm dark matter is studied in \cite{Schneider2}-
\cite{Parimbelli}.
We note, in Figure 4 of \cite{MacInnis}, or Figure 18 of \cite{Despali},
or Figure 4 of \cite{Parimbelli},
that, if the $\Lambda$CDM power spectrum is cut-off by
warm dark matter free-streaming, non-linear regeneration of $P(k)$ starts
with first galaxies, and is a \textit{major} first-order effect that should not be neglected!
One of the origial reasons for considering warm dark matter is to reduce
the counts of small galaxies with respect to the cold dark matter 
prediction that exceeds observations. This is the ``missing satellites" problem.
The suppression factor is obtained with warm dark matter simulations in
\cite{Schneider2}. Translating Equation (28) of \cite{Schneider2} to our notation,
we obtain
\begin{equation}
\frac{n_{\Lambda\textrm{WDM}}(M_\textrm{PS})}{n_{\Lambda\textrm{CDM}}(M_\textrm{PS}))} = 
\left( 1 + 0.61 \frac{M_\textrm{fs}}{M_\textrm{PS})} \right)^{-1.16}.
\label{Sch}
\end{equation}
From the \textit{data} in Figures 1, 2 and 4 of \cite{JWST} 
(for $z = 4, 5$ or 6 that have sufficient data),
we estimate the ratio of galaxy counts
$n_\textrm{data}(M_\textrm{PS})/n_{\Lambda\textrm{CDM}}(M_\textrm{PS}) \approx 0.01$
at $M_\textrm{PS} = 10^9$ M$_\odot$ (corresponding to Figure \ref{10_9}).
From (\ref{Sch}) we obtain the estimate $M_\textrm{fs} \approx 9 \times 10^{10} M_\odot$,
corresponding to $k_\textrm{fs}	\approx	2$ Mpc$^{-1}$, $v_{h\textrm{rms}}(1) \approx 170$ m/s,
$m_h \approx 0.4$ keV, and $a_{h\textrm{NR}} \approx 6 \times 10^{-7}$.
This estimate of $a_{h\textrm{NR}}$ is a factor $\approx 10$ above the limit (\ref{small-halo}) and 
a factor $\approx 2.3$ below the measurement (\ref{dwarf}).

Figure 3 of \cite{Liu} summarizes published limits on $m_h$ from several observables.
The most stringent limit for each observable is presented in Table \ref{limits}.
These limits are obtained at $z \lesssim 8$ and so are sensitive to the
regenerated power spectrum.

\begin{table}
\begin{center}
\caption{\label{limits}
Tightest published lower limits on the ``standard thermal relic" 
warm dark matter mass $m_h$ obtained from different
observables (from Figure 3 of \cite{Liu}; see citations therein).
Also shown is the corresponding upper limit on $a_{h\textrm{NR}}$,
and an approximate redshift of the measurements.
}
\begin{tabular}{lccc}
\hline
\hline
Observble                         & $m_h              $ & $a_{h\textrm{NR}}$            & Typical $z$ \\
\hline
Milky Way satellites              & $\gtrsim 10$ keV  & $\lesssim 2.3 \times 10^{-8}$   & 0 \\
Strong lensing                    & $\gtrsim 6.0$ keV & $\lesssim 3.9 \times 10^{-8}$   & 0 to 1 \\
Lyman-$\alpha$ forest             & $\gtrsim 5.2$ keV & $\lesssim 4.5 \times 10^{-8}$   & 6 \\
Galaxy UV luminosity distribution & $\gtrsim 3.2$ keV & $\lesssim 7.3 \times 10^{-8}$   & 6 to 8 \\
$\gamma$ ray burst                & $\gtrsim 1.8$ keV & $\lesssim 1.3 \times 10^{-7}$   & 4 to 8 \\
\hline
\hline
\end{tabular}
\end{center}
\end{table}

Consider the compilation of measurements of $P(k)$ in Figure 2 of \cite{MacInnis}.
The primordial linear power spectrum $P(k)$ at $z_\textrm{dec} = 1/1090$ is measured by the Planck
mission for comoving wavevectors up to $k \lesssim 0.2$ Mpc$^{-1}$, \textit{i.e.} does not reach
warm dark matter free-streaming.
All other measurements of $P(k)$ correspond to the regenerated power
spectrum. A \textit{direct} measurement of the linear $P(k)$ up to $k \approx 20$ Mpc$^{-1}$,
before the first galaxies, will become possible with
weak gravitational lensing of the cosmic microwave background (CMB) \cite{MacInnis}.

The observed reionization of the universe requires a delayed galaxy formation
compared to the $\Lambda$CDM scenario. This delay requires
$m_h = 0.51^{+0.22}_{-0.12}$ keV \cite{Lin3},
or $k_\textrm{fs} \approx 2 \pm 1$ \cite{UV}, or $m_h = 1.3^{+0.3}_{-0.7}$ keV \cite{Lapi}, or 
$m_h = 0.66^{+0.07}_{-0.08}$ keV \cite{Lapi}.
These measurements are in tension with the limits in Table \ref{limits}
and can be compared with the measurements summarized in \cite{summary}.

Small galaxies can develop due to the non-linear regeneration of small-scale
structures, and also due to ``stripping", \textit{i.e.} loss of matter to neighbors.

\begin{figure}
\begin{center}
\scalebox{0.3}
{\includegraphics{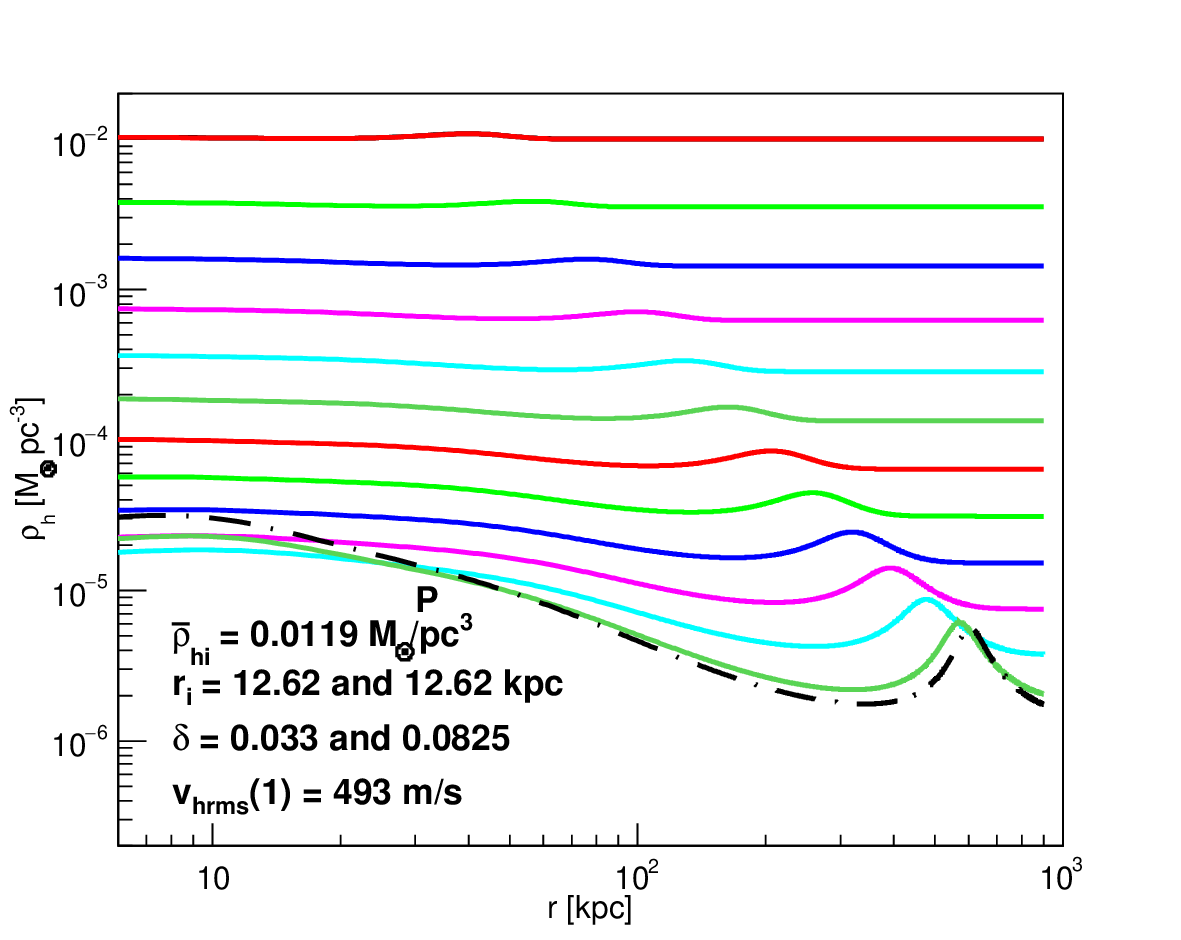}}
\scalebox{0.3}
{\includegraphics{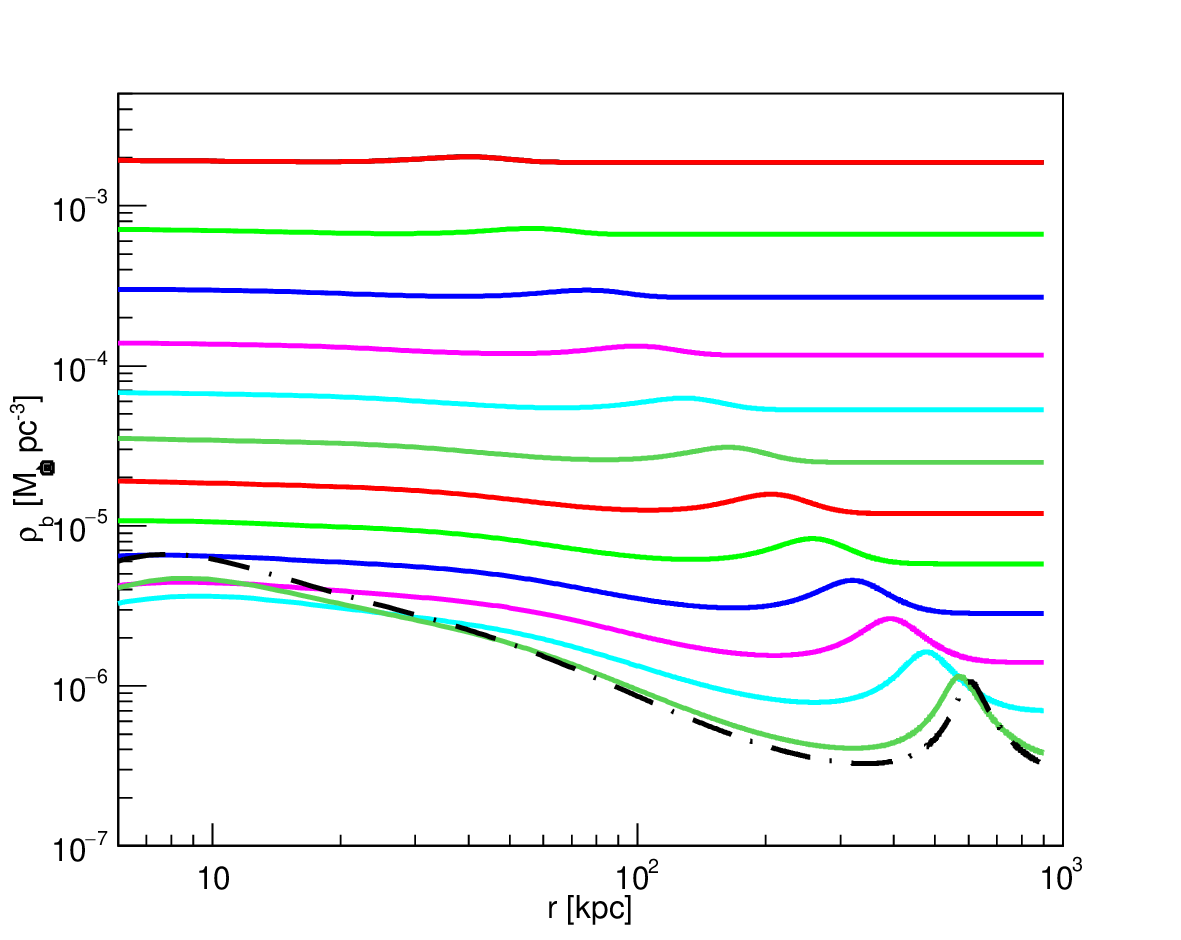}}
\scalebox{0.3}
{\includegraphics{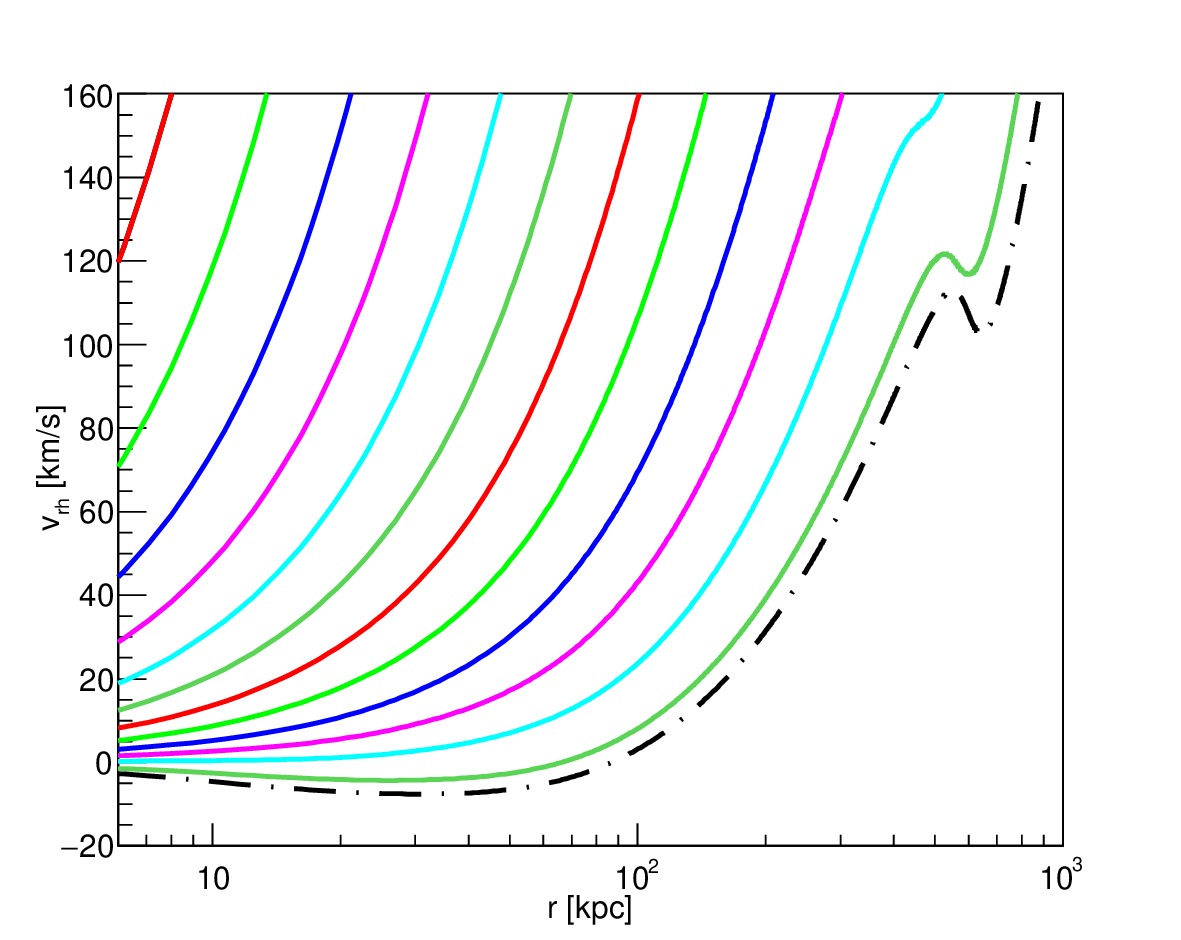}}
\scalebox{0.3}
{\includegraphics{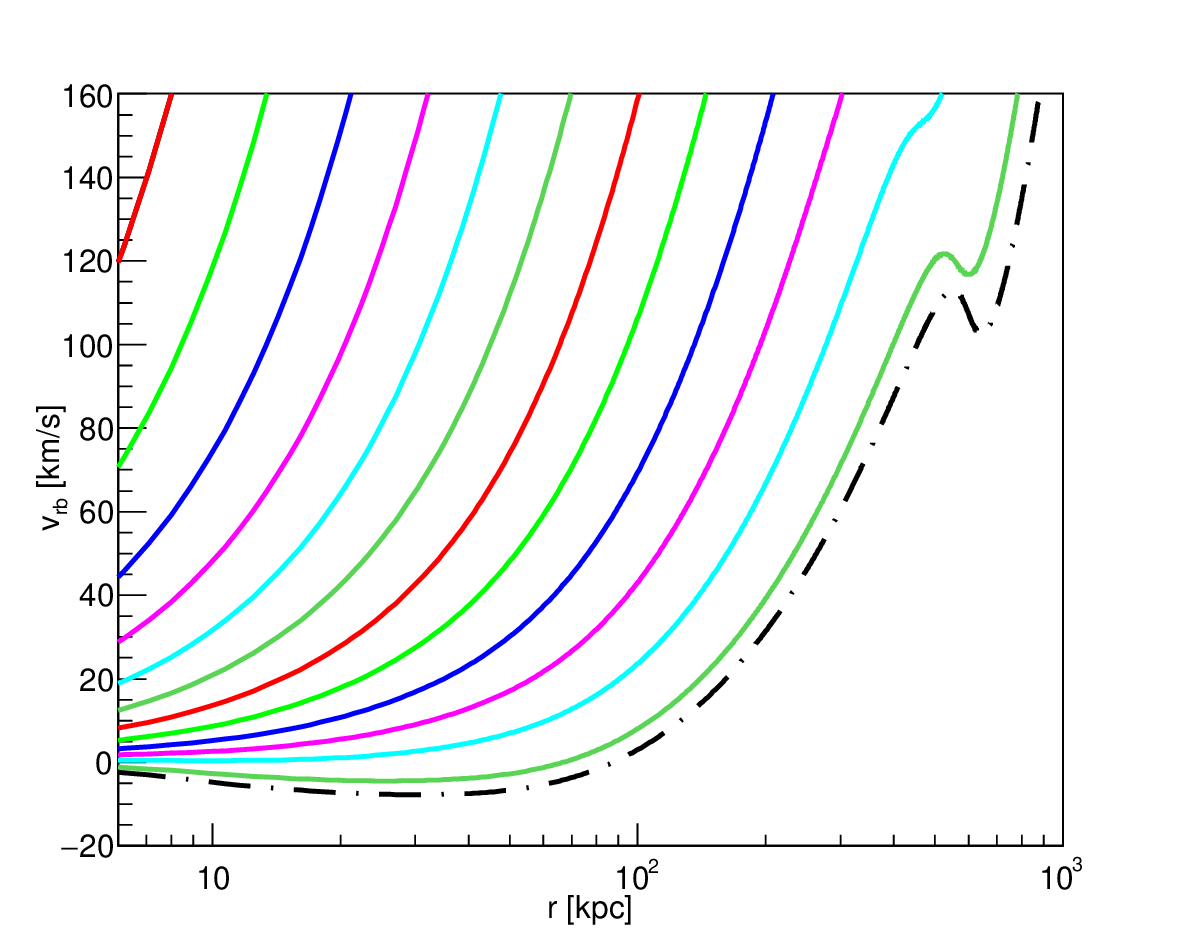}}
\caption{
The dark matter and baryon densities and radial velocities
are shown as a 
function of the proper radius $r$, at times
that increase by factors $\sqrt{2}$ (except the last dashed line).
The initial redshift of the numerical integration is $z_i = 65.9$.
The parameters of this simulation of a ``stripped-down" galaxy are
$\bar{\rho}_{hi}=0.0119$ M$_\odot/$pc$^3$,
$r_i = 12.625$ and $12.625$ kpc,
$\delta=0.033$, and $0.08254$,
$v_{h\textrm{rms}}(1)=493$ m/s.
The pivot point P has $r_{h\textrm{P}} = 32$ kpc and 
$\rho_{h\textrm{P}} = 1.3 \times 10^{-5}$ M$_\odot/$pc$^3$. 
Initial $M_\textrm{PS} = 10^{11}$ M$_\odot$. $M_h = 8 \times 10^{9}$ M$_\odot$.
$\sigma(M_\textrm{PS}, z_i, k_\textrm{fs}) = 0.036$.
$V_\textrm{rot} = 27$ km/s.
}
\label{10_11_sd}
\end{center}
\end{figure}

\section{Bottom-up and top-down evolution of galaxies}
\label{up_down}

Linear relative density perturbations are both positive and negative.
Relative to a homogenenous universe, regions with underdensity grow
faster,	and regions with overdensity grow slower, and peaks turn
around and collapse into halos. The result is underdense regions,
surrounded by sheets, that meet at filaments, that meet a nodes
where large galaxy clusters form. Galaxies punctuate the nodes and
filaments, and to a lesser extent, the sheets and voids.

Let us try to understand the formation of small galaxy satellites.
In the warm dark matter scenario, the first galaxies have
masses determined by the warm dark matter free-streaming
length, \textit{i.e.} a distribution of linear masses $M_\textrm{PS}$ of order
$M_\textrm{fs}$. The dark matter density $\rho_h(r)$ of these halos becomes equal
to the mean dark matter density of the universe at a radius
$r_\textrm{max}$ that grows in proportion to $a^{3/2}$,
faster than the separation between galaxies $\propto a$ \cite{extended}.
Therefore, as the universe expands, groups of galaxies 
overlap and coalesce, becoming larger galaxies.
Smaller galaxies become caught up between larger galaxies,
loosing matter to their expanding neighbors, and are either
absorbed completely, or collapse as ``stripped-down"
satellites with large cores (and continue loosing matter to their neighbors,
perhaps leaving behind a globular cluster).
Thus, in the warm dark matter scenario, the first galaxies
have masses of order $M_\textrm{fs}$, and, from there on,
the structure forms bottom-up and top-down, as confirmed by
simulations \cite{Paduroiu}.

Figure 1 of \cite{simulations} can help us understand
stripped-down galaxies.
In the simulation of Figure 5 of \cite{simulations},
$M_\textrm{fs} = 3.6 \times 10^{11}$ M$_\odot$	and
$m_h = 0.112$ keV, similar to our example (\ref{mh_vhrms1}).
We find that the number of
stripped-down satellites
per unit volume and a decade of mass, for a linear mass
$M_\textrm{PS} \ll M_\textrm{fs}$,
is reduced to $\approx15\%$ of the corresponding density in the
$\Lambda$CMD cosmology (some, however, do not collapse to a halo). 
Note that the Milky Way satellites are not formed as in Figure \ref{10_9},
but are stripped-down galaxies. The distribution of mass of
stripped-down galaxies extends all the way to zero, and does not exclude
the measurement (\ref{dwarf}).

The formation of a stripped-down galaxy is shown in Figure \ref{10_11_sd}.
This galaxy starts out as in Figure \ref{10_11} with $M_\textrm{PS} = 10^{11}$ M$_\odot$,
but looses matter to neighbors (with spherical symmetry	for convenience).
The pivot point	$P$ has	$r_{h\textrm{P}} = 32$ kpc, and	
$\rho_{h\textrm{P}} = 1.3 \times 10^{-5}$ M$_\odot /$pc$^3$.
So $\rho_{h\textrm{P}} r_{h\textrm{P}}^2 = 1.33 \times 10^4$ M$_\odot/$pc, compared to
$3.06\times 10^4$ M$_\odot/$pc for the galaxy with no stripping in Figure \ref{10_11}.
The parameters scale as	
\begin{equation}
\rho_{h\textrm{P}} r_{h\textrm{P}}^2 \propto r_c^{-4}, \rho_{hc}^{2/3},	M_{hc}^{4/3}, M_{200}^{2/3}, 
r_{200}^2, \left< v_{hr}^2 \right>, V_\textrm{circ}^2.
\label{stripped_down}
\end{equation}
So, with respect to the galaxy with no stripping in Figure \ref{10_11},	
the core mass $M_{hc}$ is reduced by a factor 0.54, $M_{200} \equiv M_h$ by a factor 0.29,
and the circular velocity of a test particle is reduced by a factor 0.66.

A simple empirical way to account for both non-linear regeneration
and stripped-down galaxies is to take
\begin{eqnarray}
\tau^2(k) & = & \exp{\left( -\frac{k^2}{k^2_\textrm{fs}} \right)} \qquad
\textrm{ if } k < k_\textrm{fs}, \textrm{ or } \nonumber \\
& = & \exp{\left( -\frac{k^n}{k^n_\textrm{fs}} \right)}\qquad
\textrm{   if } k \ge k_\textrm{fs},
\label{tail}
\end{eqnarray}
where $n$ is \textit{measured} to be in the range 0.5 to 1.1 \cite{UV}.
This ``tail" is sufficient to bring predictions in line with observations,
see \cite{JWST}.
The effect of $n$ on the predicted stellar mass distributions and
on the ultra-violet luminosity distributions is presented in Figure 3 of \cite{UV}.

Even the \textit{linear} power spectrum	$P(k)$ has a tail 
calculated in \cite{Boyanovsky}. For 
our example in Section \ref{discussion}, with
$k = 8.6$ Mpc$^{-1}$ and
$k_\textrm{fs} = 0.67$ Mpc$^{-1}$, we estimate $n \approx 1.08$ from
Figure 10 of \cite{Boyanovsky} for boson dark matter
that decouples while ultra-relativistic
(this is because the enhancement of low momentum bosons behaves like cold dark matter). 
In comparison, $\tau^2(k)$ of (\ref{tau2}) has a tail corresponding to $n \approx 1.36$.

\section{Baryons}
\label{baryons}

Protons recombine with electrons to form neutral hydrogen at decoupling at time $t_\textrm{dec}$.
The temperature of these ``baryons" at decoupling is $T_b = T_{\gamma 0} / a_\textrm{dec}$.
The comoving root-mean-square thermal velocity of baryons at $t_\textrm{dec}$ is
$v_{b\textrm{rms}}(1) = \sqrt{3 k_B T_b / m_b} a_\textrm{dec} \approx 3 \times 10^{-8}$ m/s.
So, after decoupling baryons, \textit{i.e.} mostly neutral hydrogen and helium atoms, 
behave as cold dark matter.

We are interested in the relative density perturbation at the onset of galaxy
formation, \textit{i.e.} at, say, expansion parameter $a_\textrm{gal} \approx 1/20$.

Consider cold dark matter plus baryons.
Let $\epsilon \equiv \delta \rho_{kh} / \bar{\rho_h} = \delta \rho_{kb} / \bar{\rho_b}$
at the time $t_\textrm{eq}$ of equal matter and radiation densities, when relative density
perturbations of dark matter start growing.
The relative density perturbation at $a_\textrm{gal}$ and at comoving wavevector $k$,
for the growing mode, is
\begin{equation}
\frac{\delta_{kh} + \delta_{kb}}{\bar{\rho}_h + \bar{\rho}_b} \approx \epsilon 
\frac{a_\textrm{gal}}{a_\textrm{eq}}.
\label{delta_cb}
\end{equation}
This is an approximation because baryon relative density perturbations start 
growing at $t_\textrm{dec}$.

Now consider warm dark matter plus baryons at $k \gg k_\textrm{fs}$ and at $a_\textrm{gal}$.
We set $\delta_{kh} = 0$. Then, for the growing mode,
\begin{equation}
\frac{\delta_{kb}}{\bar{\rho}_h + \bar{\rho}_b} \approx \frac{\epsilon \Omega_b}{\Omega_c + \Omega_b}
\left( \frac{a_\textrm{gal}}{a_\textrm{dec}} \right)^{0.3}.
\label{delta_wb}
\end{equation}
Note that the growth exponent is reduced due to $\Omega_c$,
\footnote{The calculation
is lengthy and will be omitted since replacing 0.3 by 0 only changes $n$ from 0.96 to 1.03.}
and that growth starts at $a_\textrm{dec}$.

The relative density perturbation cut-off factor
(\ref{tail}), at $a_\textrm{gal}$ and $k \gg k_\textrm{fs}$, is
obtained from the ratio of (\ref{delta_wb}) to (\ref{delta_cb}):
\begin{equation}
\tau(k) \approx \exp{\left( -\frac{1}{2} \frac{k^n}{k_\textrm{fs}^n} \right)}
\approx \frac{\Omega_b}{\Omega_c + \Omega_b} 
\left( \frac{a_\textrm{gal}}{a_\textrm{dec}} \right)^{0.3}
\left( \frac{a_\textrm{eq}}{a_\textrm{gal}} \right).
\label{tau3}
\end{equation}
The case of interest is $k = 8.6$ Mpc$^{-1}$ and $k_\textrm{fs} = 0.67$ Mpc$^{-1}$.
From (\ref{tau3}) we obtain $n \approx 0.96$. A numerical integration obtains
\begin{equation}
n \approx 0.86,
\label{tau4}
\end{equation}
in agreement with the \textit{measurement} $0.5 \lesssim n \lesssim 1.1$ \cite{UV}.
Note that the ``tail" of $\tau^2(k)$
is due, in part, to the baryons that act as cold dark matter.
If warm dark matter particles are bosons that decouple while ultra-relativistic
with zero chemical potential, then $n$ is reduced further, see figure 10 of \cite{Boyanovsky}. 

In conclusion, the distributions of
$M_\textrm{PS}$, $M_h$,	$M_*$ and $L_\textrm{UV}$
are understood down to $M_\textrm{PS} \approx 5 \times 10^8$ M$_\odot$,
or $M_h \approx 2 \times 10^8$ M$_\odot$,
over a wide redshift range, see \cite{JWST}.

This calculation leads to the following	insight.
Consider the Press-Schechter formalism
(perhaps figure 1 of \cite{simulations} may help).
A galaxy forms when the density perturbation of the sum of warm dark matter modes
up to $k \approx k_\textrm{fs}$, plus the baryon modes up to the wavevector 
$k_b > k_\textrm{fs}$ of interest,
reaches 1.69 times the mean total density.
Neglecting baryons, the	first galaxies to form have huge linear masses $M_\textrm{PS}$
of order $M_\textrm{fs}$. 
Including baryons, first galaxies form bottom-up, starting at
dwarf halos, all the way up to huge galaxies of linear mass of order $M_\textrm{fs}$,
and then continue forming hierarchically bottom-up (due to halos that overlap and coaless),
and top-down (due to halos that are stripped by their neighbors).

\section{Conclusions}
\label{conclusions}

Warm dark matter free-streaming	attenuates small-scale linear density perturbations.
However, non-linear regeneration of small-scale structure    
is a major effect. 
Furthermore, a majority of galaxies with linear mass $M_\textrm{PS} \ll M_\textrm{fs}$
are stripped-down galaxies that have lost matter to neighbors.
Finally, baryons alone are sufficient to bring the predicted galaxy stellar mass
distributions into agreement with observations.
The limit (\ref{small-halo}), based on the observation
of small Milky	Way satellite galaxies,	does not take regeneration,
nor stripping, nor baryons,
into account. With the warm dark matter
simulations in \cite{Schneider2}, an \textit{estimate} of this limit becomes
$a_{h\textrm{NR}} \lesssim 6 \times 10^{-7}$, which is a factor 
$\approx 10$ above the limit (\ref{small-halo}), and a factor $\approx 2.3$
below the measurement (\ref{dwarf}).
An empirical way to deal with non-linear regeneration, stripping, and baryons,
is to add to $\tau^2(k)$ a ``tail" as in (\ref{tail}). 
For the \textit{measured} $n$ \cite{UV},
or \textit{calculated} $n$ in equation (\ref{tau4}), we obtain
agreement with observed galaxy stellar mass distributions,
and observed galaxy ultra-violet luminosity distributions, 
over a wide redshift range, down to $M_\textrm{PS} = 5 \times 10^8$ M$_\odot$, 
or $M_h = 2 \times 10^8$ M$_\odot$, see \cite{JWST}.
So the minimum halo mass (\ref{Mh}) is not excluded.
Note that the mass distribution of stripped-down galaxies extends all the way to zero.

We are still far from a detailed and quantitative understanding 
of galaxy formation with warm dark matter. Each limit and measurement
has its own issues, so caution and an 
open mind are called for. In any case, we are unable to rule out
the \textit{measurement} (\ref{dwarf}) beyond a reasonable doubt, in spite
of all the published \textit{limits} to the contrary.

All limits on $k_\text{fs}$ that depend on galaxies need to include
in the analysis the
regeneration of small-scale structure,
the formation of stripped-down galaxies, and 
the contribution from baryons.
Limits on $k_\text{fs}$ from the Lyman-$\alpha$ forest need, in addition, 
to understand clouds of ``left over" neutral hydrogen traces in the 
re-ionized universe \cite{Keating}.
On the other hand, measurements of $v_{h\textrm{rms}}(1)$ with galaxy
rotation curves 
are direct, but have the issue of relaxation that is constrained by observations, 
see \cite{summary} and references therein.

\section*{Acknowledgments}

I thank Karsten M\"{u}ller for his early interest in this work and for many useful discussions.

\end{document}